\documentclass[12pt,preprint]{aastex}

\shortauthors{L. \ Bouchet et al.}
\shorttitle
{ \textit{INTEGRAL} SPI all-sky view}

\begin{document}

\title{\textit{INTEGRAL} SPI all-sky view in soft $\gamma$ -rays:  
Study  of point source and Galactic diffuse emissions\footnote{
Based on observations with INTEGRAL, an ESA project with instruments and science data centre
funded by ESA member states (especially the PI countries: Denmark, France, Germany, Italy, 
Spain, and Switzerland), Czech Republic and Poland with participation of Russia and USA.} }

\author{L. Bouchet, E.Jourdain, J.P.Roques}
\affil{CESR--CNRS, 9 Av. du Colonel Roche, 31028 Toulouse Cedex~04, France}

\author{A. Strong, R. Diehl} 
\affil{Max-Planck-Institut f\"ur extraterrestrische Physik, Postfach 1603, 85740 Garching, Germany}
\and

\author{F. Lebrun, R. Terrier}
\affil{DSM/DAPNIA/SAp, CEA-Saclay, 91191 Gif-sur-Yvette, France} 
\affil{APC, UMR 7164, 10 rue A. Domon et L. Duquet, 75205 Paris Cedex 13, France} 

\author{\it Received  ; accepted  }

%\altaffiltext{2}{{\it To whom proofs and offprint requests should be sent}
%\email{bouchet@cesr.fr}}

\begin{abstract}

We have processed the data accumulated with \textit{INTEGRAL} SPI during 4 years ($\sim$ 51 Ms) 
to  study the  Galactic ``diffuse'' emission morphology in the 20 keV to 8 MeV energy range. 
To achieve this objective, we have derived simultaneously an all-sky census of  emitting sources and images
of the Galactic Ridge (GR) emission.
In the central radian, the  resolved point source emission amounts to 88\%, 91\% and 68\%  of the  total emission in the
25-50, 50-100 and 100-300 keV domains respectively.
We have compared the  GR emission spatial distribution to those obtained from CO and NIR maps, 
and quantified our results through latitude and longitude profiles. Below 50 keV, the SPI data
are better traced by the latter, supporting a stellar origin for this emission.
Furthermore, we found that the GR emission spectrum follows a power law with  a photon index $\sim$
1.55 above 50 keV while an additional component is required below this energy. This component
shows a cutoff around 30 keV, reinforcing a stellar origin, as proposed by Krivonos et al. (2007).\\
%that this emission is produced by an unresolved population of accreting magnetic white dwarfs, 
%polar caps and intermediate polar caps.
%To investigate further the role of undetected sources, we have built Log~N - Log~F curves in the 
%Galactic Ridge region and estimate their potential contribution to what we ascribe to the "diffuse" emission.
The  annihilation diffuse emission component   is extracted simultaneously,
leading to the determination of the related parameters (positronium flux and fraction).
A specific discussion is devoted to the annihilation line distribution since a significant emission 
is detected over a region as large as $\sim 80^\circ$ by  $\sim 10^\circ$  potentially
associated with the disk or halo surrounding the central regions of our Galaxy.

\end{abstract}      


%%%%%%%%%%%%%%%%%%%%%%%%%%%%%%%%%%%%%%%%%%%%%%%%%%%%%%%%%%%%%%%%%%%%%%%%%
\section{Introduction}
%%%%%%%%%%%%%%%%%%%%%%%%%%%%%%%%%%%%%%%%%%%%%%%%%%%%%%%%%%%%%%%%%%%%%%%%%

The soft $\gamma$-ray GR Emission ($>$ 20 keV) has  been previously studied essentially with 
the CGRO and GRANAT missions (Purcell et al., 1996, Skibo et al.,  1997, Kinzer, Purcell \& Kurfess, 1999). 
The main conclusion was that point sources explain at least 50 \% of  the total emission up to
 $\sim$  ~200 keV, allowing the possibility to anticipate that unresolved sources 
could explain a major part of what was seen as ``diffuse continuum emission''.
Soon after its launch, first INTEGRAL results indicated    
that the GR diffuse emission is less than 15 \% of the total in the 20-200 keV domain (Lebrun  et al., 2004,
Terrier et al., 2004, Strong et al., 2005, Bouchet et al., 2005). 
In  more recent works, Revnivtsev et al. (2006), with RXTE~PCA data, and Krivonos et al. (2007), using the data
collected with the \textit{INTEGRAL} IBIS  telescope, suggest that the GRXE (Galactic Ridge X-ray Emission)
between 3 and  $\sim$ 60 keV is explained in terms of sources belonging to the population of accreting white dwarfs
binaries. \\
In a previous paper using one year of SPI data (Bouchet et al., 2005, hereafter paper I),  
we had used Galactic tracer morphologies 
to extract the spectrum of the diffuse continuum emission as well as that of the positronium/annihilation 
in the central radian of our Galaxy. We thus derived the relative contributions of these emissions along with
that of the point source emission to the total Galactic emission. In this paper, 
we  present a global view of the soft $\gamma$-ray sky emission based on a larger amount of data  
 covering the whole sky and perform  an imaging analysis of the diffuse component,
  with an estimate of its spectrum between 20 keV and 8 MeV.
 
%%%%%%%%%%%%%%%%%%%%%%%%%%%%%%%%%%%%%%%%%%%%%%%%%%%%%%%%%%%%%%%%%%%%%%%%%
\section{Instrument and Observations}
%%%%%%%%%%%%%%%%%%%%%%%%%%%%%%%%%%%%%%%%%%%%%%%%%%%%%%%%%%%%%%%%%%%%%%%%%

The ESA's INTEGRAL (INTErnational Gamma-Ray Astrophysics Laboratory) observatory 
was launched from Baikonour, Kazakhstan, on 2002 October 17.
The spectrometer SPI (Vedrenne et al., 2003) observes the sky in  the 20 keV - 8 MeV energy range 
with an energy resolution ranging from  2 to 8 keV. It consists in an array of 19   
high purity Germanium (Ge) detectors operating  at 85 K. Its geometrical surface is 508 cm$^{2}$ 
with a thickness of 7 cm. In addition to its spectroscopic capability, SPI can 
image the sky with a spatial resolution of $2.6^\circ$ (FWHM ) over a field 
of view of $30^\circ$, thanks to a coded mask located 1.7 m above the detection plane.
Despite such a modest  angular resolution, it is possible to locate intense sources with 
an accuracy of few arc minutes (Dubath et al., 2005). 
The assembly is  surrounded by a 5-cm thick BGO shield which stops and measures the flux of particles arriving
from outside the FoV.
The instrument in-fligh performance is given in Roques et al. (2003). \\
Due to the small number of detectors, SPI imaging capability relies on a specific observational strategy,
based on a dithering procedure (Jensen et al., 2003): 
the pointing direction varies around the target by steps of $2^\circ$ within 
a $5 \times 5$ square or a 7-point hexagonal pattern. In general, a pointing lasts 40 minutes,
and along its 3-day orbit, INTEGRAL can be operated $\sim 85\%$ of the time outside the radiation belts.

\begin{figure}%1

\plotone{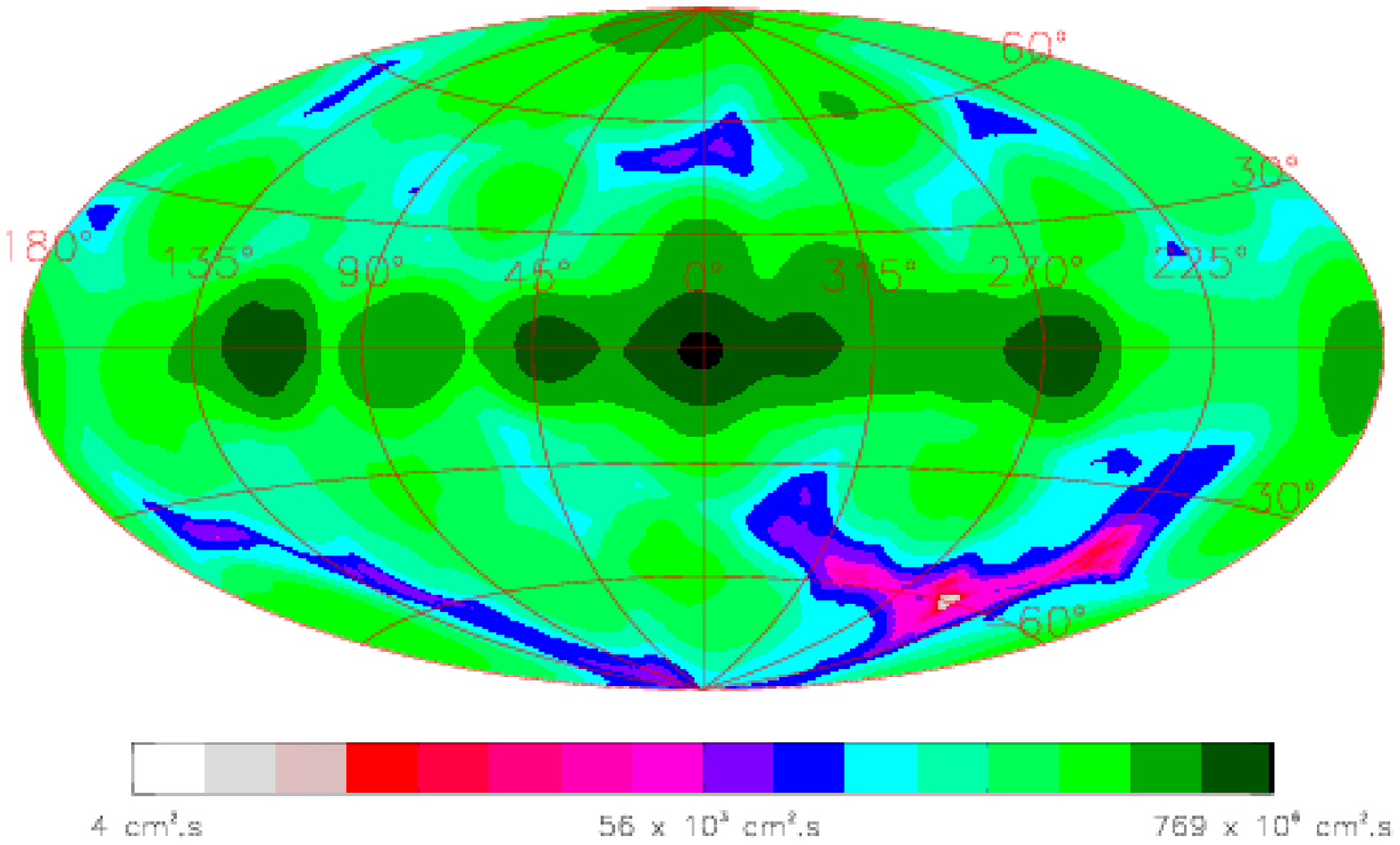}
\caption{25-50 keV \textit{INTEGRAL} SPI exposure map. Units are in cm$^2$ s. This map takes into account the differential sensitivity of SPI accross its
field of view.}
\label{fig:expo}
\end{figure}

We have analysed observations recorded from 2002, February to 2006 May, 
covering the entire sky (Fig.~\ref{fig:expo}). The central region of the
Galaxy,  within  $-50^\circ$ $\leq$ l $\leq$ $50^\circ$, $-25^\circ$ $\leq$ b $\leq$ 
$25^\circ$, is thus particularly well scanned.

Data polluted by solar flares and radiation belt entries are excluded.
After image analysis and cleaning, we obtained
51$\times$10$^6$ seconds of effective observing time.

%%%%%%%%%%%%%%%%%%%%%%%%%%%%%%%%%%%%%%%%%%%%%%%%%%%%%%%%%%%%%%%%%%%%%%%%%
\section{Technical considerations and algorithm descriptions}
%%%%%%%%%%%%%%%%%%%%%%%%%%%%%%%%%%%%%%%%%%%%%%%%%%%%%%%%%%%%%%%%%%%%%%%%%

The analysis method used in this paper is similar to the one described in our 'paper I',
with some significant enhancements however.
 We first summarize the analysis method  basics, consisting of an iterative procedure between
 source position determination (with SPIROS) and flux extraction (with Time-Model-Fit) steps.

- Source catalogues are iteratively produced  using the SPIROS software 
(Skinner \& Connell, 2003), in several 
energy bands.\\
An iterative search is performed for point sources, starting from the strongest one identified,
and successively adding weaker ones in successive iterations. Each time a source is found in 
a search of trial positions on the sky, it is added to a working source catalogue, and then used
as baseline knowledge for the next source search in the following iteration. \\
In order to avoid unnecessary artifacts from spatial-resolution limits of our instrument, for 
each newly-found source an identification step is introduced, and, for sufficient proximity, the 
(more accurate) position of the identified source is used instead. \\
In crowded regions, due to the
 2.6$^\circ$ SPI angular resolution, the brightest sources  are considered as
 representative of the local emissions. 
 For the central degree of the Galaxy 
  good fits have been obtained  in paper I, considering 1E 1740-2942.7 only.
   The better statistic of our new set of data
 requires a more complex description of this region. Thus two other (softer) sources
 have been considered below 200 keV, tentatively identified with SLX 1744-299 and IGR J17475-2822.\\

- Using the SPIROS output catalogues, source fluxes and diffuse emission intensity distribution
are derived with Time-Model-Fit (see Paper I) algorithm.\\
Time-model-fit algorithm is a model fitting procedure based on the 
likelihood/$\chi^2$ statistics. 
We thus model the sky successively with a set of three models for the large-scale diffuse and
extended emission plus the sources accumulated in the search, allowing for intensity variations
down to the time scale of individual pointings ($\sim$ 2000 sec) for the brightest point sources as well as 
for background. This latter high variability time scale is not needed normally, in particular 
our background rates are usually quite stable; but for cases such as solar flare events this
 flexibility has proven useful.
For each energy band, the Time-model-fit algorithm determines
the (time dependent) flux normalisation for point sources, extended emissions and
background. Diffuse components and strong variable sources 
contributions are then subtracted from the data to produce a "corrected" data set
used as SPIROS input for the next iteration.\\
With such flexibility, and simultaneous analysis of more than four years of data, the number 
of free parameters in such fitting approach can easily grow out of hand. Major 
improvements have been carried out, therefore: 
\begin{itemize}
\item The matrix associated with the imaging transfer function is sparse with only a few percents
of non-zero elements. We take advantage of this property by minimizing the storage and 
speeding-up the numerous matrix-vector operations through a Compressed Column Storage scheme (Duff et al. 1989). 
In addition, the transfer function related to a variable source  exhibits a peculiar structure which allows 
a direct inversion of a part of the matrix. Thanks to these optimisations the algorithm is able to
handle many more observations together with more free parameters e.g. for source variability.
\item The background determination has been refined: 
We decompose background variability to global intensity variation and into the pattern of rates
 among the 19 detectors of our camera. We retain freedom of variability for global background 
intensity as described above, and allow the general background count rate pattern in the SPI 
camera to vary on a much more restrictive time scale (on the order of months). We find that
over 6 months of data, no pattern variations are noticeable. With this compromise 
of allowing high variability for intensity parameters of model components, and modest 
variability allowance for the background pattern, we obtain adequate fits to our measurements
with a minimum number of parameters. 
\item The total energy redistribution matrix is included  for the final flux extraction,
with the proper spectral shape taken into account for each source and diffuse component.
\item The algorithm is able to handle a sky described by cells of various sizes whose intensities
are determined together with the other parameters. The use of "large" cells is suitable to
gather flux from diffuse emissions in a model independent way.
\end{itemize}

This hybrid reconstruction algorithm takes the best of methods to map point source and extended emissions
simultaneously .
It is the basis for the results presented in this paper.\\

%%%%%%%%%%%%%%%%%%%%%%%%%%%%%%%%%%%%%%%%%%%%%%%%%%%%%%%%%%%%%%%%%%%%%%%%%
\section{Mapping the sky}
%%%%%%%%%%%%%%%%%%%%%%%%%%%%%%%%%%%%%%%%%%%%%%%%%%%%%%%%%%%%%%%%%%%%%%%%%
The GR emission is difficult to measure since
its  surface brightness and its signal-to-noise ratio are low.
Moreover, as the distribution of a population of weak unresolved
point sources formally mimics an extended structure, the goal of any diffuse emission study
 consists in estimating always better the source component
in order  to derive  upper limits for the GR emission.\\
In paper I, the small amount of data, covering only the central radian of the Galaxy, justified the 
global model-fitting approach to estimate the different contributions : 
The GR Emission was tentatively described with a CO map while the annihilation process (511 keV line and
positronium continuum) was modeled by an azimuthally symmetric Gaussian of $8^\circ$ FWHM (following previous
works by Kn\"odlseder et al., 2005).\\
In this paper, we attempt to estimate a model-independent GR emission morphology, by extracting
local fluxes without spatial model.
 For this, the sky will be  represented by "large" cells of fixed sizes (tested values range from $2^\circ$ to $16^\circ$)
whose intensities are fitted to the data through the likelihood maximisation method, together with the ponctual
sources ones. \\
This represents the best  trade off between  a priori (model dependent) information introduced 
in the  model-fitting algorithm (positions of a catalog of known sources plus variability timescale of each source)
and model independent results (fluxes of sources and of cells of the extended emission).
\\
The first step consists thus to build a catalog of individual sources as complete as possible, following the 
procedure used in Paper I. Then, large cells are added in the convergence process to obtain the 
diffuse emission distribution.

%%%%%%%%%%%%%%%%%%%%%%%%%%%%%%%%%%%%%%%%%%%%%%%%%%%%%%%%%%%%%%%%%%%%%%%%%
\subsection{Source catalogue}
%%%%%%%%%%%%%%%%%%%%%%%%%%%%%%%%%%%%%%%%%%%%%%%%%%%%%%%%%%%%%%%%%%%%%%%%%
We used the method described in paper I to generate source catalogues: the SPIROS software delivered in the 
INTEGRAL OSA (Off-line Scientific Analysis) package is used iteratively in conjunction with our hybrid algorithm.
The latter  calculates  variable source flux contributions using  Galactic tracers models
for spatial morphologies of the interstellar emissions (8$^\circ$ axisymmetric Gaussian for the annihilation emission,
 DIRBE 4.9~$\mu$ 
and CO  maps for the GR continuum below and above 120 keV respectively),for which
a normalisation factor is adjusted during the fitting process. The extended emission and variable point source
 contributions are then removed from the data set provided to SPIROS for the next iteration.\\
 It should be noted that the point source fluxes remain the same (within error bars) whatever the GR continuum
 distribution model is used (CO or DIRBE), as expected given the low surface brightness of this component.

As explained in  paper I, to minimize the error bars (maximize the signal-to-noise ratios), we need to 
describe the sky with a minimum number of free parameters. We thus define a sky model for  
each of the energy bands.
In practice, 12 sources (marked with '*' in table I)  have been considered as variable for the
25-50 keV band. Above 50 keV, the reduced $\chi^2$ being sufficiently close to 1, only Cyg X-1 is set variable up to 200 keV.\\

\begin{figure}%2

\includegraphics[width=18 cm,height=22cm]{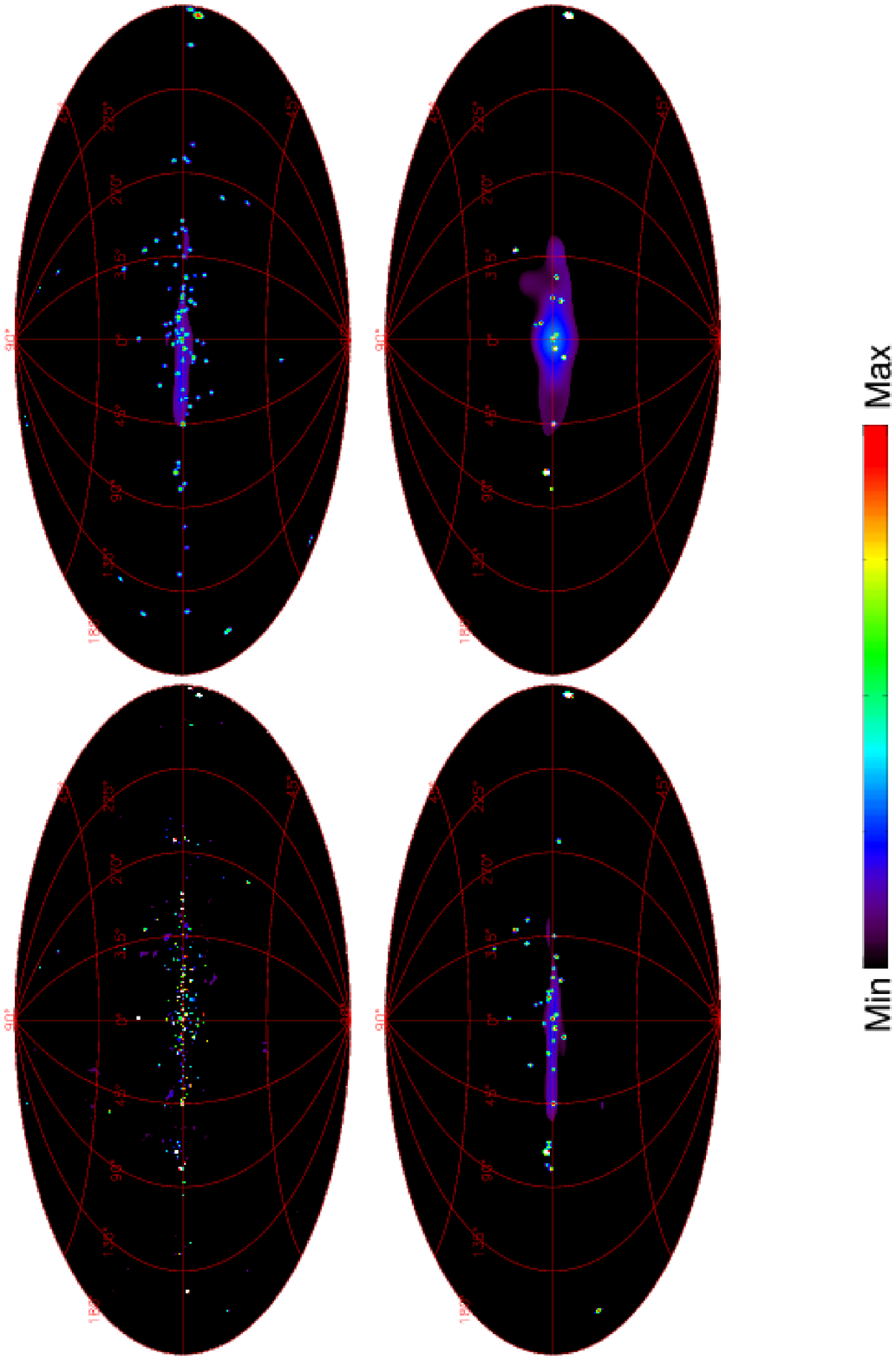}
\caption{Sky-maps in the 25-50 keV (top-left), 50-100 keV (top-right),
100-200 keV (bottom-left) and 200-600 (bottom-right) energy bands.
The images are scaled logarithmically with a rainbow color map 
(the scale of colors ranges from black (weakest intensity) to red (strongest intensity)). 
The scale is saturated to reveal the weakest sources. There are some systematics 
due to strong variable sources such as Cyg X-1 combined with the finite
precision of SPI response, especially in the 25-50 keV band. We take into 
account this systematic by using a higher threshold in this domain.}
\label{fig:maps}
\end{figure}

Figure~\ref{fig:maps} displays the resulting sky images sum of point  sources and ``diffuse'' components in different energy 
bands and illustrates their  evolution. At low energy, the sky emission is dominated by 
sources  while a "diffuse/extended" structure appears above 200-300 keV, in a domain
corresponding  to the annihilation radiation.
%, as illustrated in the  4th panel of Figure~\ref{fig:maps}.
%This image    has been built in a narrow band around 511 keV without the a priori 8$^\circ$ axisymmetric
% Gaussian distribution in order to reveal the emission spatial distribution straightly from the data. In
%this 10 keV wide band, source contribution is less than 10\% while the annihilation
% processus lights up the central regions of our Galaxy.
 We will come back to it in the next section, dedicated to the diffuse emissions.
Above 511 keV, sources again dominate the sky emission.

The resulting catalogues contain 173, 79, 30 and 12 sources detected 
above $\sim 3.5 \sigma $ in the 25-50 keV and the  50-100 keV bands and above $\sim 2.5 \sigma$ in the 100-200 keV and 200-600 keV 
ones
(see Table I).
All of them except one are associated (within 1$^\circ$) with at least one IBIS source (Bird et al., 2006).

Above 600 keV only the Crab Nebula, Cyg X-1, GRS1915+115 and GRS1758-258 are detected,  the 2 former being
still emitting above 2 MeV.

Note that the reported fluxes are  four years averaged values, and that this analysis is not 
optimised for any particular source, since all data are adjusted simultaneously. An individual source analysis should be
based on a restricted number of exposures, selected on the basis of the  pointing direction (typically less
than $\sim 12^\circ$ relatively to the source direction), and requires a detailed study of the source
time evolution. 

%XXXXXXXXXXXXXXXXXXXXXXXXXXXXXXXXXXXXXXXXXXXXXXXXXXXXXXXXXXXXX
% VERIFIER SI C'EST UTILE
  % (mis dans les improvements)

%Once the final catalog realised, a last step has been included to take into account the off-diagonal
%terms of the energy redistribution matrix (position search and flux extraction have been
%performed using only the photopeak efficiency).  
%For each source, the  complete projected pattern  (taken into account Aperture Response Function and total energy 
%Redistribution Matrix Function) has been calculated and a last run performed to obtain final
%fluxes.

%XXXXXXXXXXXXXXXXXXXXXXXXXXXXXXXXXXXXXXXXXXXXXXXXXXXXXXXXXXXXXXXXXXXXX

%%%%%%%%%%%%%%%%%%%%%%%%%%%%%%%%%%%%%%%%%%%%%%%%%%%%%%%%%%%%%%%%%%%%%%%%%
\subsection{GR Diffuse emission}
%%%%%%%%%%%%%%%%%%%%%%%%%%%%%%%%%%%%%%%%%%%%%%%%%%%%%%%%%%%%%%%%%%%%%%%%%
 
%%%%%%%%%%%%%%%%%%%%%%%%%%%%%%%%%%%%%%%%%%%%%%%%%%%%%%%%%%%%%%%%%%%%%%%%%
\subsubsection{morphology}
%%%%%%%%%%%%%%%%%%%%%%%%%%%%%%%%%%%%%%%%%%%%%%%%%%%%%%%%%%%%%%%%%%%%%%%%%
Once the contribution of the individual sources has been  independently estimated, we can 
study the unresolved component.
The a priori information will thus now be introduced in the source terms.
To determine the spatial distribution of the Galactic Ridge emission, we have considered the region 
$|l|<100 ^\circ$, $|b|<20^\circ$ and divided it into cells of size  $\delta l = 16^\circ$ x
  $\delta b = 2.6 ^\circ$, the 511 keV line case being treated apart.
These cell or pixel sizes have been chosen a posteriori to optimize the signal-to-noise
ratio per cell while being  sufficiently small to follow the observed spatial variations.
We use the a priori information on the source positions obtained in the previous step, whereas 
the ``diffuse'' pixel cells, sources and background intensities are to be fitted to the data.
The number of unknowns is high but reasonable 
compared to the data  and the problem  is easily tractable by a simple likelihood optimisation to
 determine all the corresponding intensities and error bars.

\begin{figure}%3
\includegraphics[width=14 cm,height=18 cm]{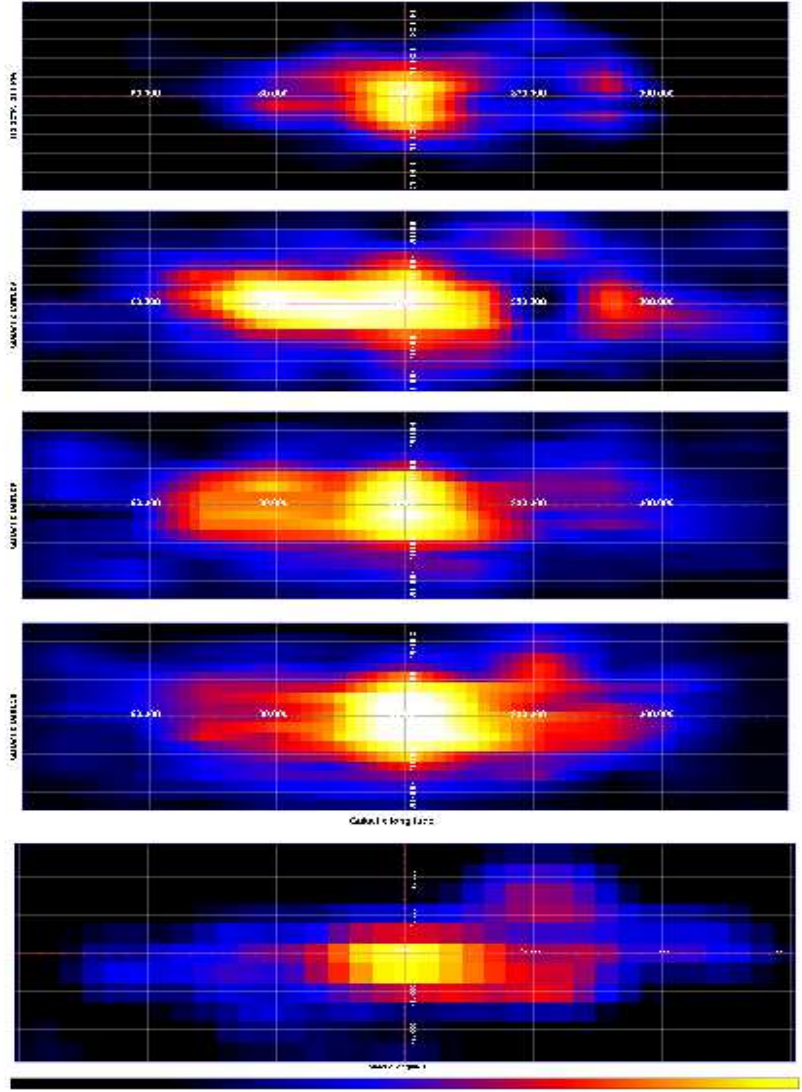}
\caption{Diffuse emission morphology in different energy bands (significance maps):
25-50 keV, 50-100 keV, 100-200 keV, 200-600 keV and 1.8-7.8 MeV  from up to bottom.
%Note that for the last image  (E $>$1.8 MeV), the latitude scale is larger.
}
\label{fig:diffuse}
 
\end{figure}

Figure~\ref{fig:diffuse} displays the  images obtained through this method for the  ``diffuse''
component(s) in different energy bands. The last one contains the $^{26}$Al 1.8 MeV line
but it is clear that its contribution to the large band flux is quite negligible.
 %The spatial extension seems to evolve  with energy,
 %particularly above 1 MeV, where the broader distribution could be related to an emission mecanism
 %of different nature.

\begin{figure}%4
\includegraphics[width=8.8cm,height=17.0cm]{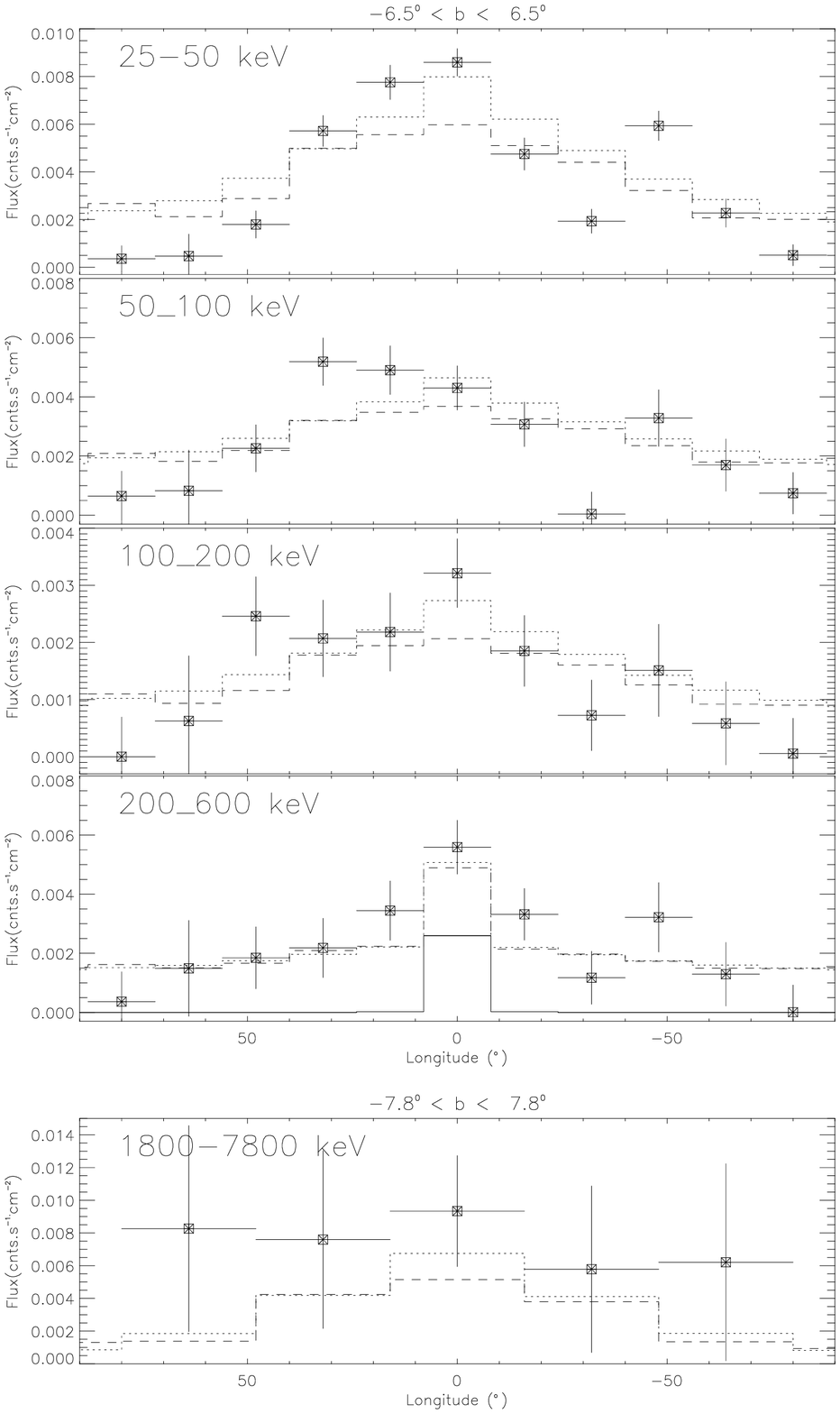}
\caption{Longitude profiles in different energy bands for $|$b$|$ $\leq$ $6.5^\circ$ 
(except for the last band).
Dotted and dashed lines correspond respectively to the NIR 4.9 $\mu$ and CO maps.
For the 200-600 keV band, the annihilation contribution ($8^circ$gaussian distribution)
is represented by the solide line.}
\label{fig:profilong}
\end{figure}

\begin{figure}%5
\includegraphics[width=8.8cm,height=17.0cm]{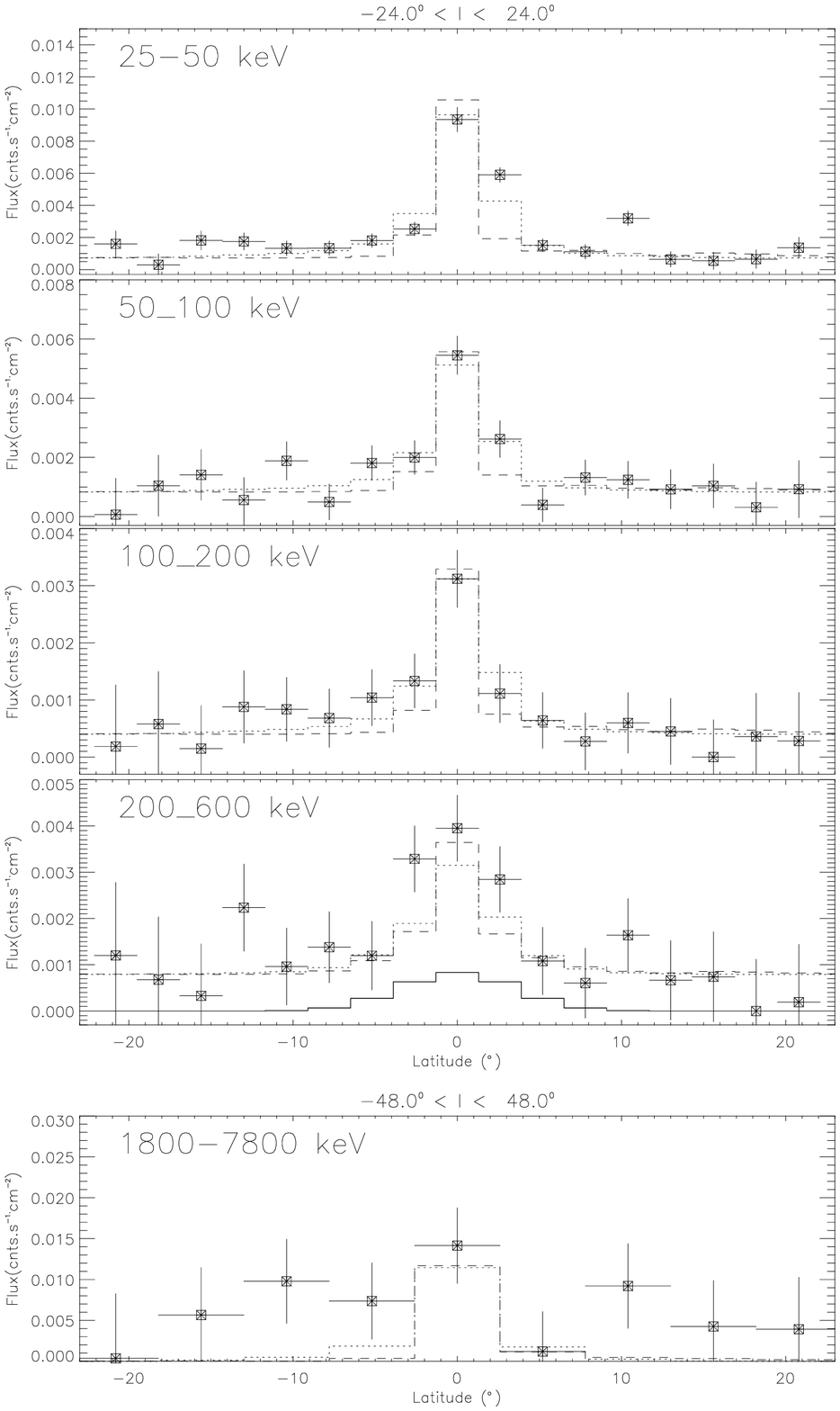}
\caption{Latitude profiles in different energy bands for $|$l$|$ $\leq$ $24^\circ$
(except for the last band).
Dotted and dashed  lines correspond respectively to the NIR 4.9 $\mu$ and CO maps.
For the 200-600 keV band, the annihilation contribution ($8^circ$gaussian distribution)
is represented by the solide line.}
\label{fig:profilat}
\end{figure}

To quantify more easily the behavior of the diffuse emission, we present our results in terms of 
longitude and latitude profiles (figures~\ref{fig:profilong} and ~\ref{fig:profilat}). They have been built 
by integrating the flux measured for  $|b|<6.5^\circ$ or $|l|<24 ^\circ$, in   longitude
and latitude bins  respectively. We can then compare them to those obtained from CO (Dame et al. 2001) and
 NIR Dirbe 4.9~$\mu$ corrected from reddening (http://lambda.gsfc.nasa.gov) maps.

Both models agree grossly with SPI longitude profiles, showing a slowly decreasing intensity
 of the GR emission toward high longitudes.
\\
The latitude profiles are not so similar:
Up to $\sim$ 100 keV, the GR high energy and NIR emissions
 seem to be concentrated in a region slightly more 
extended than the CO one. This is supported by calculating a $\chi^2$ between the resulting sky images 
and the proposed models. We obtained  $\chi^2$ values 
of 74 and 86 (37 dof) in the 25-50 keV energy band, while in the 50-100
keV energy band, chisquare values are of 50 and 53 for DIRBE and CO maps, respectively. 
 This fits in with the current understanding of 
a stellar origin for the GR emission in the X ray domain ($<50$ keV) proposed by  Krivonos et al. (2007).
However, at energies below ~50 keV, systematic effects from instrumental properties and/or variability 
of the sources provides limitations to fit quality.
\\
    
The signal to noise ratio has a mininum in the 600~keV-1.5~MeV domain, prohibiting any serious analysis
 but at high  energies ($>$ 1.8 MeV), the SPI data analysis shows clearly that the diffuse emission is detected up to
several MeV. All the compared spatial distributions are here compatible but
the expectation is that the high energy GR diffuse emission coincides with the CO regions, where CR
interact.\\

\begin{figure}%6
\includegraphics[width=18 cm,height=5 cm]{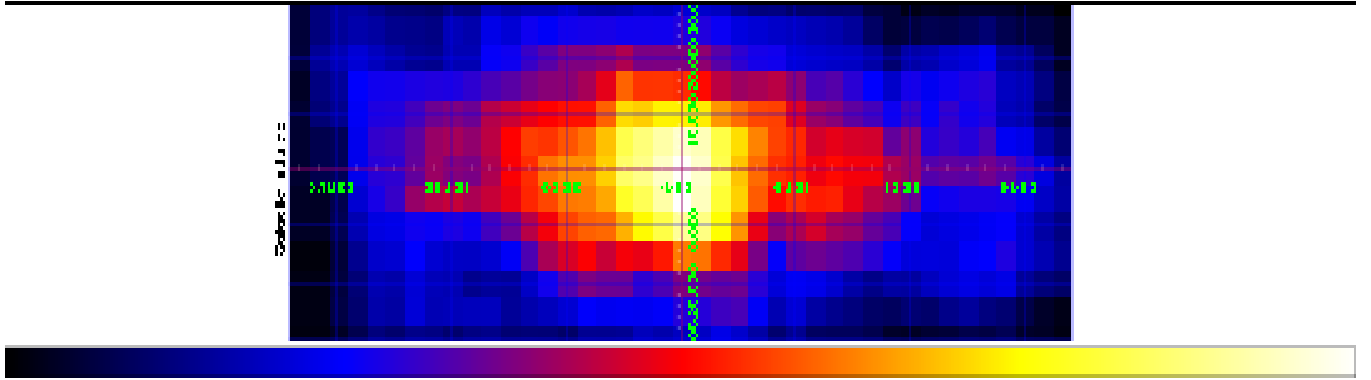}

\caption{map of the significance of the 511 keV line emission 
}
\label{fig:511map}
\end{figure}

\begin{figure}%7
\includegraphics[width=8.8cm,height=4.0cm]{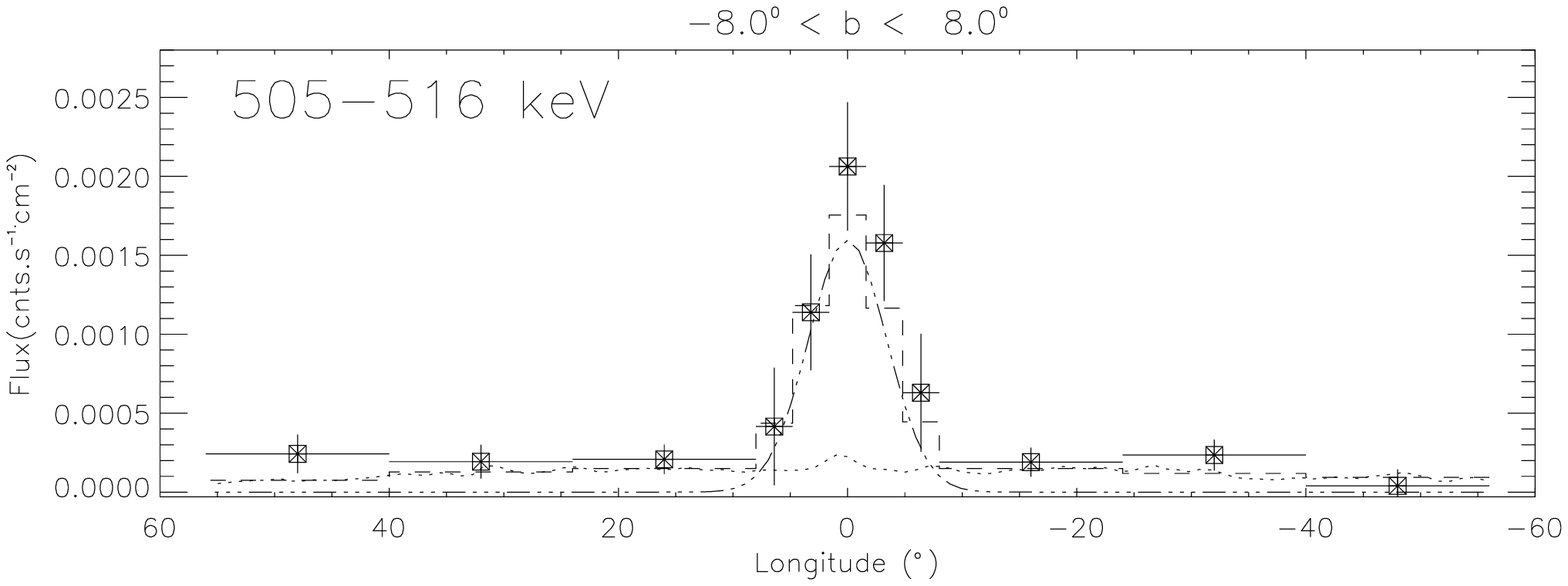}
\caption{Longitude profile in the 511 keV line for $|$b$|$ $\leq$ $8^\circ$.
Dotted-dashed  line corresponds to a $8^\circ$ axisymetric Gaussian. Dotted line corresponds
to the extended distribution (240 $\mu$ map model). The sum of both has been integrated 
on the same bins as the data (histogram) to compare to them.}
\label{fig:profilong511}
\end{figure}

\begin{figure}%8
\includegraphics[width=8.8cm,height=4.0cm]{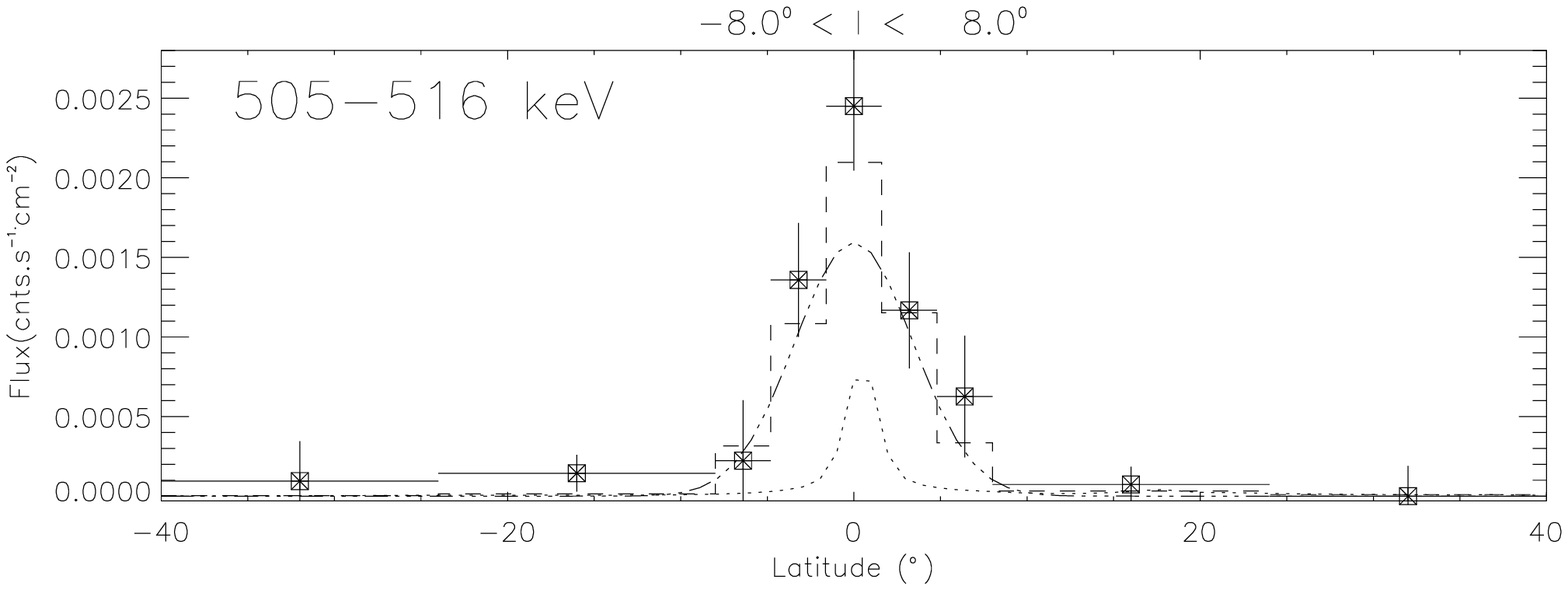}

\caption{Latitude profile in the 511 keV line  for $|$l$|$ $\leq$ $8^\circ$.
 Dotted-dashed  line corresponds to a $8^\circ$ axisymetric Gaussian. Dotted line corresponds
to the extended distribution (240 $\mu$ map model). The sum of both has been integrated 
on the same bins as the data (histogram) to compare to them.}
\label{fig:profilat511}
\end{figure}

A  deeper analysis has been performed for the annihilation line emission morphology.
A 505-516 keV sky map has been built using cells of size
 $\delta l = 5^\circ$ x  $\delta b = 5^\circ$ (fig.~\ref{fig:511map}). 
In this narrow energy range, the emission is almost entirely due to
the annihilation line, the contribution of the other components being negligible. The emission
profiles (figures~\ref{fig:profilong511} and ~\ref{fig:profilat511}) are peaked towards
 the Galactic Centre. This  structure, corresponding to the bulge, has been modeled with an 
  axisymmetric Gaussian of  8.0$^\circ \pm 0.9^\circ$ 
 in good agreement with previous \textit{INTEGRAL} SPI
results (e. g. Kn\"odlseder et al. (2005)).
 However, the map together with the profiles suggest that the emission is not limited 
 to that structure but
exhibits  an additional  extented component revealed by the long exposure.

To quantify  this result, we attempted to describe the extended spatial distribution superimposed
to the central bulge, with various models. Simple geomerical shapes (ie two-dimensional 
Gaussians) have been  tested as well as maps obtained in other wavelengthes (CO and   DIRBE  
 maps). The best results are obtained with NIR data (more precisely the 3.5 to
240~$\mu$ maps, too close  to be differentiated in these tests) which, independently, happen 
to be good tracers of the $^{26}Al$ line emission (Kn\"odlseder et al., 1999).\\
 For a sky model consisting of the
 240~$\mu$ map plus a 8$^\circ$ axisymmetric Gaussian, we obtain  fluxes of
$1.7\pm 0.3 \times 10^{-3} photons~cm^{-2}~s^{-1}$, in the extended structure 
and  $ 0.87 \pm 0.06 \times 10^{-3}  photons~cm^{-2}~s^{-1}$ in the central one.
Even though it is difficult to describe the emission in more detail, this represents a good indication 
for a bulge/disk structure.\\
% Coming back to the central bulge emission, we also tested several possibilities
% (one or two Gaussians
%of varying widths,  centered on (l=0$^\circ$, b=0$^\circ$))
%and come to the conclusion that 
%the best adjustment is obtained with two Gaussians 
%of  FWHM 3.0 $^\circ$ and 10.4 $^\circ$ with fluxes  of $ 2.0 \pm 0.5 \times 10^{-4}  photons~cm^{-2}~s^{-1}$ and
%$ 7.7 \pm 0.7 \times 10^{-4}  photons~cm^{-2}~s^{-1}$ respectively.
%In this configuration, the extended emission (with the 240~$\mu$ map distribution) gathers
%$ 1.7 \pm 0.33 \times 10^{-3}  photons~cm^{-2}~s^{-1}$.
% We point out here the probable
%complexity of the bulge geometry and a precise study of its morphology would require
%a specific analysis.
Coming back to the central bulge emission, we also tested several possibilities
as we suspect a geometry more complex than a single axisymetric gaussian.
A precise study of its morphology would require a specific analysis, but first investigations
show that the fluxes corresponding to each structure vary within the error bars.

%%%%%%%%%%%%%%%%%%%%%%%%%%%%%%%%%%%%%%%%%%%%%%%%%%%%%%%%%%%%%%%%%%%%%%%%%
\subsubsection{Central radian Spectrum}
%%%%%%%%%%%%%%%%%%%%%%%%%%%%%%%%%%%%%%%%%%%%%%%%%%%%%%%%%%%%%%%%%%%%%%%%%

\begin{figure}%9
\plotone{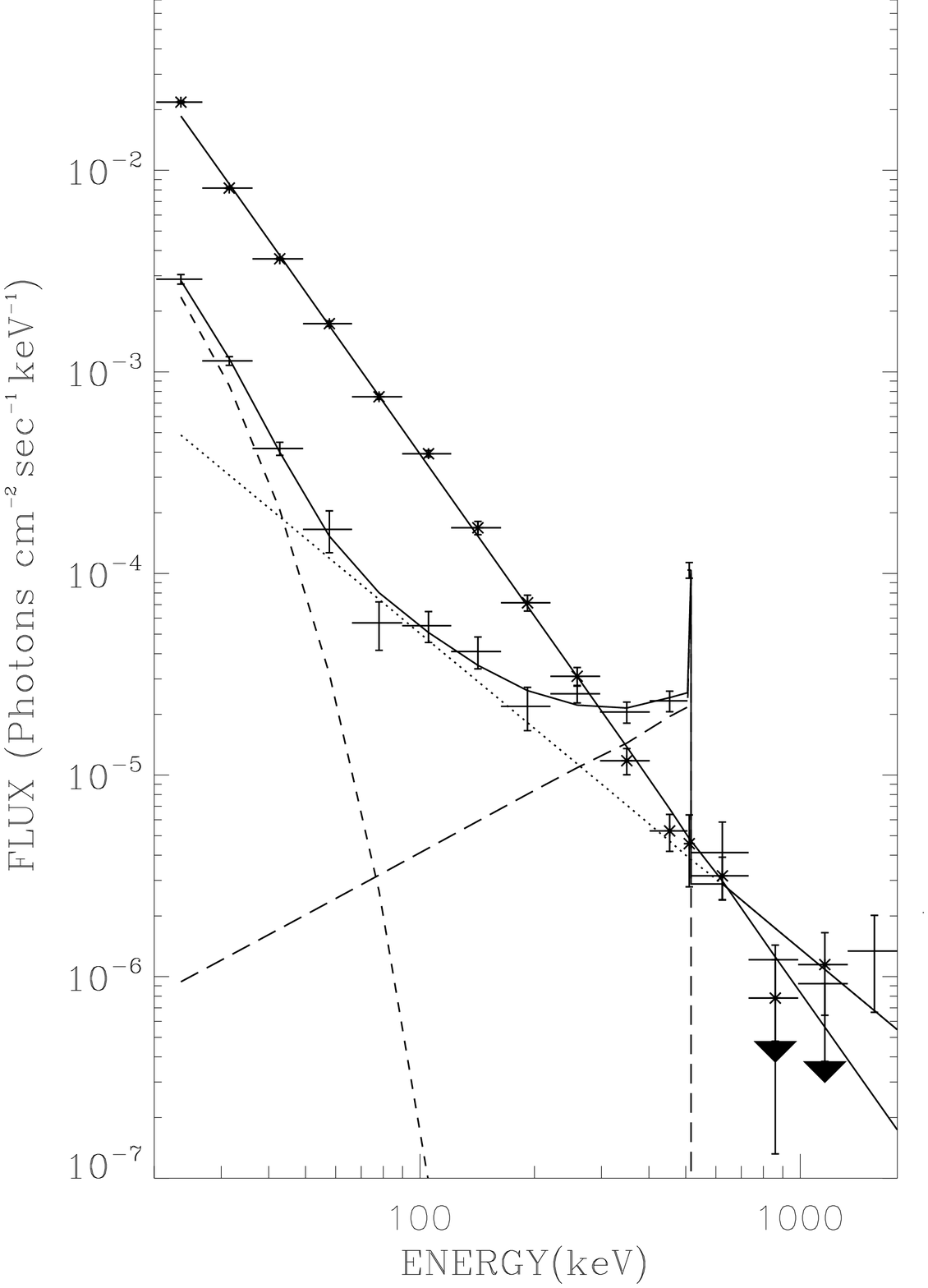} 
\caption{Spectra of the different emission components in the central radian of the Galaxy, 
for $|$l$|$ $\leq$ $30^\circ$ and $|$b$|$ $\leq$ $15^\circ$: Sum of  sources (stars),
 annihilation spectrum (long dashed line) and
total ``diffuse'' emission (solid line). ``Diffuse'' contiunuum components are described by a power law (dotted line) 
and power law plus exponential cutoff (dashed line).}
\label{fig:spectrum}
\end{figure}

The spectral shape of the diffuse continuum remains of prime importance to determine its origin.
As this emission is concentrated in the central regions, the spectral analysis has been restricted
to the Galactic central radian   ($|l| < 30^\circ$ and $|b| < 15^\circ$).\\
The ``diffuse'' continuum spectrum shown in fig~\ref{fig:spectrum} has been extracted assuming a 
NIR 4.9~$\mu$ spatial distribution (up to 120 keV) and a CO map (above 120 keV) and fitted with 3 components.
\begin{itemize}
\item The diffuse spectrum (apart from positronium) is fitted by a power law of index $1.55 \pm 0.25$, with
a 100 keV flux of $4.8 \pm 0.6 \times  10^{-5} ~ photons~cm^{-2}~s^{-1}~keV^{-1}$.
\item The additional component below $\sim$ 50 keV presents a curved shape, which can be modeled by a 
power law of  index fixed to 0 with an exponential cutoff at $7.5 \pm 1 $ keV and a 50 keV flux of 
$6.6 \pm 0.5 \times  10^{-5} ~ photons~cm^{-2}~s^{-1}~keV^{-1}$. These parameters differ  from those
derived in Paper I, since many more sources have been identified, and thus removed
from 'diffuse" component. This remaining "diffuse" low energy component, with
a central radian luminosity of $\sim~ 1 \times 10^{37} erg.s^{-1}$, is likely to correspond to the population
of accreting magnetic white dwarfs proposed to provide a dominant contribution to the Galactic
 X-ray emission (Krivonos el al.,  2007).
\item The third component is due to the positronium/annihilation emission with its characteristic 
shape. We extract it using  a  8 $^\circ$ Gaussian spatial distribution and obtained 
a 511 keV line flux of $8.68 \pm 0.61  \times  10^{-4} ~ photons~cm^{-2}~s^{-1}$.
A fit to the data allows us to determine a positronium flux of $3.6 \pm 0.42 \times 
 10^{-3} ~ photons~cm^{-2}~s^{-1}$ corresponding to  a  positronium fraction (as defined by
 Brown and Leventhal, 1987) of $F_p$ =$ 0.98 \pm 0.05$.
\item The 511 keV emission component reveals a disk component in addition to the well-known
 bright bulge.  We estimate a bulge-to-disk ratio of  $\sim 0.5$.
\end{itemize}
Above   results, synthetized in Table 2, are consistent with our results in Paper I (except for 
the newly-found disk component at 511 keV).

%%%%%%%%%%%%%%%%%%%%%%%%%%%%%%%%%%%%%%%%%%%%%%%%%%%%%%%%%%%%%%%%%%%%%%%%%
\section{Sources vs diffuse emission contribution}
%%%%%%%%%%%%%%%%%%%%%%%%%%%%%%%%%%%%%%%%%%%%%%%%%%%%%%%%%%%%%%%%%%%%%%%%%
In Fig~\ref{fig:spectrum} we compare the  different  contributions to the Galactic emission :
The  total point source emission  spectrum has been built by adding spectra from all sources
detected between $|l| < 30^\circ$ and $|b| < 15^\circ$. It can be roughly 
described  between 20 keV and 1 MeV by a power law   
  with a photon index of $2.67 $ and a flux at 50 keV of
  $2.47   \times  10^{-3} ~ photons~cm^{-2}~s^{-1}~keV^{-1}$.
The diffuse emission is represented  together with its 3 components. 

>From the analysis presented above, we can deduce  ratios of combined emission of detected sources to 
the total emission 
of 88\%, 91\% and 68\% in the 25-50, 50-100 and 100-300~keV bands respectively.
These values represent  lower limits as a population of weak unresolved sources may be 
confused with a diffuse emitting structure.

%%%%%%%%%%%%%%%%%%%%%%%%%%%%%%%%%%%%%%%%%%%%%%%%%%%%%%%%%%%%%%%%%%%%%%%%%
\section{Discussion}
%%%%%%%%%%%%%%%%%%%%%%%%%%%%%%%%%%%%%%%%%%%%%%%%%%%%%%%%%%%%%%%%%%%%%%%%%

 The unresolved/extended  emission observed in the soft $\gamma$ -ray domain
comprises three separate components: The electron-positron interactions
are known  to produce a large scale emission. In another hand,
 an interstellar emission is expected due to high energy particles travelling 
 inside the whole galaxy. The third component has been identified more recently,
with a contribution exponentially decreasing above ~10 keV. We discuss each of them 
in the following sections.

\subsection{$e{^-}/e{^+}$ interaction}
Between 300 and 511 keV, the annihilation process plays the major role.\\
We have determined a 511 keV flux  of  8.7 $\pm 0.6 \times 10^{-4}~photons~cm^{-2}~s^{-1}$
and 
a positronium fraction $F_{p}$ = $0.98 \pm 0.05$.
Comparison with previous results is quite satisfying :  for example, Kinzer et al. (2001) 
reported  a positronium fraction of  $0.93 \pm 0.04$ in OSSE data, while independent 
SPI data analyses led to  $F_{p}$ = $0.94 \pm 0.06$ (Churazov et al. 2005),
$F_{p}$ = $0.97 \pm 0.08$ (Paper I), $F_{p}$ = $0.92 \pm 0.09$ (Weidenspointner et al., 2006) and $F_{p}$ = $0.967 
\pm 0.022$ (Jean et al. 2006). Fluxes values are more difficult to compare as they depend
on the assumed spatial distribution but they range around 1~$\times 10^{-3}~photons~cm^{-2}~s^{-1}$ in the quoted references. \\
Indeed, the 511 keV line emission spatial distribution
is essentially concentrated within the central region of the Galaxy but
its morphology may be more complex than the proposed axisymetric ($\sim 8^\circ$) Gaussian distribution.
 The regions  surrounding this central part contain a
significant flux ($\sim 1.7 \pm 0.3 \times 10^{-3}  photons~cm^{-2}~s^{-1}$) which is not compatible with 
this simple spatial distribution. As a function of longitude, this emission
seems to extend symetrically up to $\sim 40-50^\circ$, while in latitude, it extends $\sim 15-20^\circ$.
%This suggests either  a more complex morphology for the 511 keV and/or an extended disk emission. In the latter case,
This emission could for example correspond to the disk component, as already suggested by the OSSE team (Kinzer et al. 1999).
Among several models, the  NIR emission distributions, otherwise considered as good $^{26}Al$ tracers, reveal themselves to give the best 
description of the SPI data in this energy domain. 
  The mentionned link with the $^{26}Al$ emission 
leads to a straightforward interpretation:  the $^{26}Al$ decay produces 1809 keV photons
 simultaneously with  positrons and thus 511 keV photons in such a way that
 
$Flux(511 keV) = 0.85 \times (2-1.5 \times fp)\times Flux(1809 keV)$\\
where fp is a positronium fraction of the annihilation process.\\
Using the COMPTEL measurement F(1809 keV) = $ 7-10\times 10^{-4}  photons~cm^{-2}~s^{-1}$ (Kn\"odlseder, 1999)
and assuming fp= 0.98, we expect a 511 keV flux of $ 3-5\times 10^{-4}  photons~cm^{-2}~s^{-1}$.
We  thus conclude that ~ 20-30\% of the 511 keV flux emitted in the disk/halo structure could
be explained by the $^{26}Al$ decay. The remaining $\sim 75\%$ of the observed disk/halo emission 
require  another origin and will be refined by adding future observations.\\ 

Previous ratio estimates of the bulge-to-disk flux ratios have been obtained 
with OSSE/CGRO,  varying from 0.2 to 3.3 depending upon whether
the bulge component features a halo (which leads to a large ratio) or not (Milne et al. 2000, Kinzer et al., 2001). 
Benefiting from a more uniform Galactic plane coverage than the OSSE data, the SPI data allow to 
better constrain this parameter.
Indeed, even though the bulge-to-disk flux ratio  depends on the assumed bulge and disk shapes,
tests with several representative configurations lead to a  bulge-to-disk flux ratio
estimation of $\sim$ 0.5.

 \subsection{The $<$ 50 keV component }

 This low energy component deserves a short discussion: it has been found to
follow the same spatial distribution as the NIR DIRBE 4.9 $\mu$ emission (see fig.~\ref{fig:profilat}).
 Together with 
the  soft spectral shape, this corroborates  the interpretation
of the Galactic ridge emission between 20 and 60 keV by Krivonos et al. (2007) in terms of a 
population of accreting magnetic white dwarfs, which present a spectral cutoff around 30-50 keV
 (Suleimanov et al, 2005). Our observed luminosity of $\sim$ 1 $\times$ $10^{37}~erg~s^{-1}$ 
between 20 and 60 keV in the central radian is in good agreement with the estimation  of 1.23 $\pm$ 0.05 
 $\times$ $10^{37}~erg~s^{-1}$ given by Krivonos et al. (2007) 
on the basis of the proposed model. However, it is obviously beyond the SPI capacities (and 
objectives) to resolve this emission since the potential objects  which could explain it
are too faint, even thought the global emission is clearly detected as a large scale distribution.\\

\subsection{Interstellar emission}

\begin{figure}%10
%\plotone{f10.ps}
\plotone{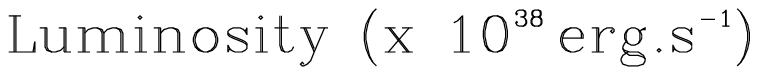}
\caption{SPI all sky diffuse emission spectrum (triangles) compared to the broad band spectrum compiled by Krivonos et al. (2007).
Small diamonds are \textit{RXTE} PCA ($<$17 keV) and \textit{INTEGRAL} IBIS ($>$17 keV) data, the big ones \textit{CGRO} OSSE \& EGRET data.
 The dotted line represent the SPI Galactic Ridge emission  best fit.}
\label{fig:spectreGR}
\end{figure}

Above 50 keV, we  detected an emission following a power law,
 which joins the high energy points previously measured by \textit{CGRO} OSSE (Kinzer et al. 1999) in the MeV region 
 (fig~\ref{fig:spectreGR}). Our photon index  of $\sim 1.55 \pm 0.25$ is quite compatible with the $\sim$ 1.75 value  
 reported by these authors, in the 200 keV to 10 MeV domain.
  Krivonos et al. (2007) do not detect any Galactic Ridge Emission 
  in the ~ 60-200 keV range.  The SPI measurements give marginal detections at those energies, 
  fully compatible with their upper limits, as illustrated on fig~\ref{fig:spectreGR}.
To go further on this point, we have  to improve the determination of the source emission contribution.
Indeed,  if the Galactic ridge emission intensity measured with our model fitting procedure depends only slightly on the 
assumed spatial morphology,  the  number of  sources  included in the sky model can lead to different results.
It is clear that a population of weak sources, below the current SPI sensitivity, could contribute to
the detected emission.
% We have thus built Log~N - Log~F curves and extrapolated  them
%toward fainter sources. Our conclusion is that if  the Galactic ridge emission is really due to a large
%number of sources whose intensities are between 1 and $\sim$~3 mCrab, we should be able to derive 
%constraining values in the framework of an  extended INTEGRAL mission (so far confirmed 
%until 2010), more particularly below 100 keV.
\\

Still, the nature of this emission  can be actually related to high energy particles.
 At the energies of interest to SPI the main gamma-ray emission process is
inverse-Compton (IC) emission from
relativistic electrons; non-thermal bremsstrahlung is of minor importance.
The electrons for IC have  typical energies of 1 GeV and below, and the
important
radiation fields are CMB, starlight and IR.
A theoretical prediction in the SPI energy range using the GALPROP model
 (Strong \& Moskalenko, 1998; Strong et al, 2004;
Moskalenko et al., 2007; Strong, Moskalenko and Ptuskin, 2007)
 can be seen in Strong et al (2005).
The general conclusion up to now has been that the IC component fails by a
factor of a few
to explain the observed emission.\\
However now in this new work, with three times as much SPI data it is
possible to determine the
non-thermal component with much greater accuracy, both because of
increased exposure
and better accounting for point sources and better handling of the
instrumental background.

 The result is a significantly lower intensity of the power-law  component
 than given by Strong et al (2005), where however the error bars were  much larger than
 now.
The result is that now there is better agreement with the IC prediction,
mitigating the need for other processes. In fact, the hard power law continuum 
is now in good agreement with predictions from the GALPROP code, which takes 
into account both cosmic-ray
 primary electrons and secondary electrons/ positrons: the hard power law 
continuum component can be identified with inverse Compton scattering.
Detailed  modelling and interpretation of the data will
be given in Porter et al. (2007).

%%%%%%%%%%%%%%%%%%%%%%%%%%%%%%%%%%%%%%%%%%%%%%%%%%%%%%%%%%%%%%%%%%%%%%%%%
\section{ Conclusions}
%%%%%%%%%%%%%%%%%%%%%%%%%%%%%%%%%%%%%%%%%%%%%%%%%%%%%%%%%%%%%%%%%%%%%%%%%
 
The "diffuse emission" generic term gathers a set of various processus depending on the energy domain,
and making the determination of its origin rather difficult.
In the soft $\gamma$-ray domain, the diffuse emission represents a complex problem, as its presence is swamped 
with the individual sources emission and difficult to disentangle.\\
To investigate it, we have  developed an imaging algorithm dedicated to extended structures. 
This code first determines
and takes into account the point sources emitting in the same energy domain, 
allowing us to then better estimate the  geometry of the regions producing diffuse emissions.\\
The SPI all-sky survey analysis reveals that 173 sources can be identified in the 25-50 keV with 30 of them
emitting a significant flux ($>$ $\sim$ 2.5 $\sigma$) above 100 keV. 
It is clear now that the sources dominate the Galactic emission up to $\sim$ 300 keV (Fig.~\ref{fig:spectrum}). 
Finally, upper limits on the diffuse emission are estimated to be one tenth of the total emission
below 100 keV and one third in the 100-300 keV band. \\

 The $e{^-}/e{^+}$ annihilation produces an emission which has already been the subject of many studies.
The main point  emerging from our analysis concerns the spatial distribution of this emission since
a large structure, potentially associated with the Galactic disk/halo  has been found to contain
 a significant fraction of the 511 keV line total flux. While a deeper analysis is required to refine 
 this result, it is clear that the  $\sim 8^\circ$  Gaussian distribution previoulsy  reported
does not represent the complexity of the  511 keV line emission morphology.\\
Once the source and annihilation process contributions have been taken into account, we can
 access to the "diffuse" Galactic emission, whose origin remains problematic.\\
Its  morphology is of prime importance as it is thought to trace the CR electron population, observed 
through
 bremsstrahlung or Compton emission. 
The \textit{INTEGRAL} SPI data allow us to investigate this particular topic,
 and we present for the first time images of the diffuse Galactic emission up to a few MeV.\\
This diffuse emission is clearly detected between 50 keV and 2 MeV with a power law relatively
hard (photon index around 1.55), and an additional component, much steeper, required below 50 keV.\\
Toward high energy, the SPI spectrum joins with the OSSE and EGRET measurements. All together, they can be
compared to various models to understand the nature of this emission. The inverse Compton 
interaction is a good candidate and will be investigated in a forthcoming paper.
Other interstellar processes have been invoked to explain the emission below 1 MeV
(Dogiel et al., 2002; Masai et al., 2002;
Schr\"oder et al., 2005).
In another side, unresolved point sources are the most probable origin below 50 keV (Krivonos et al., 2007)
 and are also a possibility at higher energies as AXPs or other pulsars (see for exemple Kuiper et al. 2004;
 Strong, 2007). 
Observations of such sources up to a few MeV would allow to quantify their potential contribution
and bring a new piece to the Galactic diffuse emission riddle.

%\begin{acknowledgements}
\section*{Acknowledgments}  The \textit{INTEGRAL} SPI project has been completed under the responsibility and leadership of CNES.
   We are grateful to ASI, CEA, CNES, DLR, ESA, INTA, NASA and OSTC for support.

 AWS is supported by the German Bundesministerium f\"ur Bildung,
  Wissenschaft, Forschung und Technologie (BMBF/DLR) under contract No.
  FKZ 50 OG 0502.
%\end{acknowledgements}

%%%%%%%%%%%%%%%%%%%%%%%%%%%%%%%%%%%%%%%%%%%%%%%%%%%%
%%%%%%%%%% Figures 
%%%%%%%%%%%%%%%%%%%%%%%%%%%%%%%%%%%%%%%%%%%%%%%%%%%%
\clearpage

\newpage

\clearpage
\clearpage

\newpage
\newpage
\newpage

\newpage

\newpage

\newpage
\newpage

%%%%%%%%%%%%%%%%%%%%%%%%%%%%%%%%%%%%%%%%%%%%%%%%%%%%%%%%%%%%%%%%%%%%%%%%%%%%%
%%%% Table 1 %%%%%%%%%%%%%%%%%%%%%%%%%%%%%%%%%%%%%%%%%%%%%%%%%%%%%%%%%%%%%%%%
%%%%%%%%%%%%%%%%%%%%%%%%%%%%%%%%%%%%%%%%%%%%%%%%%%%%%%%%%%%%%%%%%%%%%%%%%%%%%
\begin{deluxetable}{lcccccc}%1
\tablewidth{0pt}
\tablecaption{Sources catalogue }

\tabletypesize{\scriptsize}
\tablehead{
\colhead{Name}
&\colhead{l} 
&\colhead{b}   
&\colhead{25 -  50 keV} 
&\colhead{50 - 100 keV} 
&\colhead{100 - 200 keV} 
&\colhead{200 - 600 keV} \\
%&\colhead{~}         
&\colhead{deg} 
&\colhead{deg} 
&\colhead{mCrab} 
&\colhead{mCrab} 
&\colhead{mCrab} 
&\colhead{mCrab} }
\startdata
      A0535+26$^{*}$ &   181.45 &     -2.6 &          78.4 $\pm$   1.3 &          33.6 $\pm$   2.0 &                 $<$   3.9 &                           \\
                Crab &   184.55 &     -5.8 &        1000.0 $\pm$   0.8 &        1000.0 $\pm$   1.9 &        1000.0 $\pm$   3.7 &        1000.0 $\pm$   8.3 \\
         4U~0614+091 &   200.87 &     -3.4 &          21.1 $\pm$   1.3 &          13.1 $\pm$   3.1 &                 $<$   5.5 &                           \\
     IGR~J07597-3842 &   254.48 &     -4.7 &           3.0 $\pm$   0.6 &           5.6 $\pm$   1.5 &           6.4 $\pm$   2.7 &                $<$   12.0 \\
     IGR~J07565-4139 &   256.68 &     -6.7 &           3.5 $\pm$   0.5 &           3.5 $\pm$   1.5 &                 $<$   5.6 &                           \\   
         4U~0836-429 &   261.94 &     -1.1 &          19.2 $\pm$   0.4 &          13.9 $\pm$   1.1 &           6.5 $\pm$   2.2 &                $<$   10.2 \\
     IGR~J09103-3741 &   261.97 &      7.0 &           1.8 $\pm$   0.5 &           1.9 $\pm$   1.3 &                 $<$   2.5 &                           \\
       MCG-05-23-016 &   262.70 &     17.2 &           6.3 $\pm$   1.5 &           5.1 $\pm$   3.3 &                 $<$   5.2 &                           \\
      Vela~X-1$^{*}$ &   263.05 &      3.9 &         190.4 $\pm$   0.7 &          24.0 $\pm$   1.1 &           3.8 $\pm$   2.2 &                $<$   5.1 \\
         Vela~pulsar &   263.55 &     -2.8 &          10.8 $\pm$   0.4 &           8.8 $\pm$   1.1 &          12.6 $\pm$   2.1 &         13.9 $\pm$  4.9  \\  
     IGR~J09026-4812 &   268.84 &     -1.1 &           2.5 $\pm$   0.4 &                 $<$   1.1 &                           &                           \\ 
  SWIFT~J1009.3-4250 &   273.93 &     10.8 &           4.6 $\pm$   0.7 &           2.7 $\pm$   1.9 &                 $<$   3.5 &                           \\     
           4U0919-54 &   275.81 &     -3.9 &           3.0 $\pm$   0.5 &           3.2 $\pm$   1.2 &                 $<$   4.6 &                           \\
            LMC~X-4 &   276.33 &    -32.5 &          38.3 $\pm$   1.3 &           9.6 $\pm$   3.6 &                 $<$   7.0 &                           \\
            NGC~4388 &   279.16 &     74.3 &          12.6 $\pm$   1.2 &          18.4 $\pm$   3.1 &          14.7 $\pm$   5.5 &                 $<$  23.6 \\ 
        EXO~0748-676 &   279.98 &    -19.8 &          16.7 $\pm$   1.5 &          16.8 $\pm$   3.7 &                 $<$   6.3 &                           \\
       GRO~J1008-57 &   282.98 &     -1.8 &           4.5 $\pm$   0.5 &           3.5 $\pm$   1.4 &                 $<$   5.4 &                           \\
     IGR~J09025-6814 &   284.17 &    -14.2 &           3.8 $\pm$   1.0 &                 $<$   2.7 &                           &                           \\
     IGR~J10147-6354 &   286.70 &     -6.1 &           2.6 $\pm$   0.6 &                 $<$   1.6 &                           &                           \\
            ESO~33-2 &   287.84 &    -33.3 &           4.4 $\pm$   1.2 &                 $<$   3.3 &                           &                           \\
              3C~273 &   290.00 &     64.3 &           8.5 $\pm$   0.9 &          13.8 $\pm$   2.5 &           9.4 $\pm$   4.9 &                 $<$  22.2 \\
             Cen~X-3 &   292.09 &      0.3 &          30.1 $\pm$   0.6 &           3.2 $\pm$   1.5 &                 $<$   5.8 &                           \\
     IGR~J11305-6256 &   293.94 &     -1.5 &           2.2 $\pm$   0.6 &                 $<$   1.6 &                           &                           \\
   1E~1145.1-6141 &   295.49 &     -0.0 &          23.2 $\pm$   0.6 &          13.1 $\pm$   1.5 &                 $<$   2.8 &                           \\
      IGRJ12026-5349 &   295.72 &      8.4 &           3.2 $\pm$   0.7 &                 $<$   1.7 &                           &                           \\
            NGC~4593 &   297.51 &     57.4 &           3.5 $\pm$   0.9 &           3.6 $\pm$   2.6 &                $<$   10.2 &                           \\
     XSS~J12270-4859 &   298.97 &     13.8 &           3.3 $\pm$   0.9 &                 $<$   2.4 &                           &                           \\
            NGC~4507 &   299.65 &     22.9 &           6.2 $\pm$   1.3 &                 $<$   3.5 &                           &                           \\
      GX~301-2$^{*}$ &   300.10 &     -0.0 &         110.2 $\pm$   1.1 &           7.6 $\pm$   1.7 &                 $<$   3.4 &                           \\
             SMC~X-1 &   300.39 &    -43.6 &          17.3 $\pm$   3.0 &          10.4 $\pm$   7.9 &                 $<$  13.7 &                           \\             
         4U~1246-588 &   302.67 &      3.8 &           6.0 $\pm$   0.6 &           9.0 $\pm$   1.6 &           8.4 $\pm$   3.1 &               $<$   13.6 \\
     IGR~J13020-6359 &   304.11 &     -1.1 &           5.2 $\pm$   0.7 &           3.6 $\pm$   1.7 &                 $<$   6.4 &                           \\
            NGC~4945 &   305.27 &     13.3 &          17.6 $\pm$   0.8 &          18.9 $\pm$   2.1 &          14.3 $\pm$   4.0 &          20.2 $\pm$   8.3 \\
         ESO323-G077 &   306.02 &     22.3 &           3.7 $\pm$   1.1 &                 $<$   2.9 &                           &                           \\
          4U~1323-62 &   307.03 &      0.5 &          10.8 $\pm$   0.7 &           4.9 $\pm$   1.8 &                 $<$   3.3 &                           \\
               Cen~A &   309.51 &     19.4 &          38.2 $\pm$   1.0 &          42.8 $\pm$   2.5 &          56.9 $\pm$   4.6 &          70.1 $\pm$   9.8 \\
          4U~1344-60 &   309.76 &      1.5 &           6.0 $\pm$   0.7 &           6.9 $\pm$   1.6 &           4.4 $\pm$   3.1 &                 $<$  14.0 \\
     Circinus~galaxy &   311.32 &     -3.8 &          15.1 $\pm$   0.7 &          11.8 $\pm$   1.7 &                 $<$   3.3 &                           \\
IGR~J14331-6112 &   314.90 &     -0.7 &          10.2 $\pm$   0.6 &           7.0 $\pm$   1.6 &          12.3 $\pm$   2.9 &                $<$  12.8 \\
IGR~J14471-6414$^{+}$  &   315.00 &     -4.1 &           4.2 $\pm$   0.7 &                 $<$   1.8 &                           &                           \\
            IC~4329A &   317.51 &     30.9 &          12.8 $\pm$   1.1 &          14.9 $\pm$   2.9 &          11.1 $\pm$   5.3 &                 $<$  11.5 \\
%     IGR~J14175-4641 &   317.87 &     13.7 &           2.2 $\pm$   0.7 &           7.8 $\pm$   1.7 &          12.9 $\pm$   3.1 &                 $<$   6.8 \\
IGR~J14536-5522$^{+}$ &   319.76 &      3.4 &          11.3 $\pm$   0.6 &          12.5 $\pm$   1.4 &           7.9 $\pm$   2.8 &                 $<$   6.1 \\
        PSR~B1509-58 &   320.31 &     -1.2 &           3.8 $\pm$   0.6 &           3.7 $\pm$   1.6 &           9.4 $\pm$   3.0 &                           \\
          4U~1626-67 &   321.80 &    -13.1 &           9.1 $\pm$   1.0 &           4.6 $\pm$   2.6 &                 $<$   4.7 &                           \\
             Cir~X-1 &   322.12 &      0.0 &           2.1 $\pm$   0.6 &                 $<$   1.5 &                           &                           \\
     IGR~J16377-6423 &   324.53 &    -11.5 &           7.2 $\pm$   0.8 &           9.0 $\pm$   1.9 &                 $<$   7.0 &                 $<$   7.6 \\
      IGRJ16119-6036 &   325.23 &     -6.7 &           4.4 $\pm$   0.6 &           4.2 $\pm$   1.5 &                 $<$   2.8 &                           \\
       XTE~J1550-564 &   325.88 &     -1.8 &          19.3 $\pm$   0.5 &          27.6 $\pm$   1.3 &          30.2 $\pm$   2.5 &          18.8 $\pm$   5.5 \\
         4U~1538-522 &   327.43 &      2.2 &          12.8 $\pm$   0.5 &                 $<$   1.3 &                           &                           \\
% End correction page : Note that a very supurious source has been doscarded (it is not used in fact)   

          ESO~138-1  &   329.66 &     -9.5 &           3.6 $\pm$   0.6 &                 $<$   1.6 &                           &                           \\
         4U~1608-522 &   330.92 &     -0.9 &          13.9 $\pm$   0.6 &           7.0 $\pm$   1.4 &                 $<$   2.7 &                           \\
     IGR~J15479-4529 &   332.42 &      7.0 &           7.7 $\pm$   0.5 &           4.3 $\pm$   1.2 &                 $<$   4.8 &                           \\
PSR~J1617-5055$^{+}$ &   332.45 &     -0.3 &           9.9 $\pm$   0.6 &           8.0 $\pm$   1.4 &           6.2 $\pm$   2.3 &                 $<$  10.4             \\
         4U~1636-536 &   332.91 &     -4.8 &          19.3 $\pm$   0.5 &           9.1 $\pm$   1.3 &           3.7 $\pm$   2.5 &                 $<$   5.7 \\
     
     IGR~J16318-4848 &   335.61 &     -0.4 &          36.2 $\pm$   1.2 &          15.4 $\pm$   1.4 &           4.9 $\pm$   2.8 &                 $<$   6.3 \\

    4U~1630-47$^{a}$ &   336.91 &      0.2 &          34.4 $\pm$   2.1 &    26.7 $\pm$   1.4 &          21.7 $\pm$   2.4 &          21.4 $\pm$   5.3 \\                               
IGR~J16358-4726$^{a}$ &   337.10 &      0.0 &           6.8 $\pm$   2.1 &    14.5 $\pm$   5.5 &                 $<$  10.7 &                           \\

%    4U~1630-47$^{a}$ &   336.91 &      0.2 &          34.4 $\pm$   2.1 &          14.5 $\pm$   5.5 &                 $<$  10.7 &                           \\
%IGR~J16358-4726$^{a}$ &   337.10 &      0.0 &           6.8 $\pm$   2.1 &          26.7 $\pm$   1.4 &          21.7 $\pm$   2.4 &          21.4 $\pm$   5.3 \\

            GX~339-4 &   338.94 &     -4.3 &          36.7 $\pm$   0.5 &          40.1 $\pm$   1.3 &          35.6 $\pm$   2.4 &          30.0 $\pm$   5.2 \\
            GX~340+0 &   339.59 &     -0.1 &          16.6 $\pm$   0.6 &           3.0 $\pm$   1.3 &                 $<$   2.4 &                           \\
SPI J1720-49$^{***}$ &   340.46 &     -6.8 &           4.4 $\pm$   0.6 &          11.1 $\pm$   1.4 &                 $<$   2.5 &                           \\
   
    IGR~J16493-4348$^{+}$  &   341.37 &      0.6 &           7.3 $\pm$   0.6 &           2.8 $\pm$   1.3 &                 $<$   2.6 &                           \\
   
         4U~1705-440 &   343.32 &     -2.3 &          14.1 $\pm$   0.7 &           3.6 $\pm$   1.6 &           6.6 $\pm$   3.0 &                $<$   13.8 \\
         4U~1702-429 &   343.87 &     -1.3 &           9.4 $\pm$   0.7 &           7.7 $\pm$   1.7 &                 $<$   3.3 &                           \\
        OAO~1657-415 &   344.37 &      0.3 &          69.1 $\pm$   0.6 &          33.5 $\pm$   1.3 &          10.3 $\pm$   2.2 &          13.0 $\pm$   4.8 \\
        GRO~J1655-40 &   344.98 &      2.5 &           2.0 $\pm$   0.5 &           6.3 $\pm$   1.1 &           9.1 $\pm$   2.2 &                 $<$   4.9 \\
         4U~1735-444 &   346.05 &     -7.0 &           8.4 $\pm$   0.5 &                 $<$   1.3 &                           &                           \\
     IGR~J17195-4100 &   346.98 &     -2.2 &           6.7 $\pm$   0.5 &           4.8 $\pm$   1.2 &                  $<$  4.6 &                           \\
   4U~1700-377$^{*}$ &   347.75 &      2.2 &         189.6 $\pm$   0.8 &         103.1 $\pm$   1.1 &          35.5 $\pm$   2.2 &           6.4 $\pm$   4.8 \\
     IGR~J16194-2810 &   349.08 &     15.5 &           6.9 $\pm$   0.8 &                 $<$   1.8 &                           &                           \\
            GX~349+2 &   349.11 &      2.7 &           8.8 $\pm$   0.9 &                 $<$   2.2 &                           &                           \\

     IGR~J17091-3624 &   349.52 &      2.2 &           7.8 $\pm$   0.8 &          10.0 $\pm$   1.0 &           5.6 $\pm$   2.0 &           9.8 $\pm$   4.4 \\
     IGR~J16500-3307 &   349.71 &      7.3 &           2.5 $\pm$   0.6 &                 $<$   1.4 &                           &                           \\
 
  Swift~J1656.3-3302 &   350.61 &      6.3 &           3.4 $\pm$   0.5 &           8.5 $\pm$   1.0 &           5.0 $\pm$   1.9 &          17.6 $\pm$   4.0 \\
 
    IGR~J17204-3554  &   351.27 &      0.7 &           2.5 $\pm$   0.6 &                 $<$   1.5 &                           &                           \\
   
     IGR~J16482-3036 &   351.43 &      9.2 &           2.7 $\pm$   0.5 &           7.1 $\pm$   1.1 &           5.2 $\pm$   1.9 &                 $<$   4.3 \\
     
       EXO~1722-363  &   351.50 &     -0.3 &           8.9 $\pm$   0.6 &           4.3 $\pm$   1.5 &           5.1 $\pm$   2.8 &                           \\
          4U~1705-32 &   352.79 &      4.7 &           2.1 $\pm$   0.4 &           3.0 $\pm$   1.0 &                 $<$   4.0 &                           \\
         1A~1744-361 &   354.12 &     -4.2 &           3.1 $\pm$   0.6 &           3.4 $\pm$   1.4 &                 $<$   2.7 &                           \\
            GX~354-0 &   354.31 &     -0.1 &          34.2 $\pm$   0.4 &          14.6 $\pm$   0.9 &           7.6 $\pm$   1.8 &           7.9 $\pm$   4.1 \\
       XTE~J1720-318$^{+}$ &   354.62 &      3.1 &           7.6 $\pm$   0.4 &           5.0 $\pm$   0.9 &                 $<$   3.4 &                           \\
        GRS~1724-308 &   356.31 &      2.3 &          17.0 $\pm$   0.4 &           8.7 $\pm$   0.9 &                 $<$   3.4 &                           \\
       XTE~J1710-281 &   356.35 &      6.9 &           4.8 $\pm$   0.5 &           3.3 $\pm$   1.1 &                 $<$   2.2 &                           \\
        SLX~1746-331$^{+}$ &   356.81 &     -3.0 &           8.0 $\pm$   0.5 &           6.4 $\pm$   1.1 &                 $<$   4.4 &                           \\
         3A~1822-371 &   356.86 &    -11.3 &          19.7 $\pm$   0.5 &           4.7 $\pm$   1.1 &                 $<$   4.4 &                           \\
     
  IGR~J17464-3213$^{*}$ &   357.26 &     -1.8 &          17.5 $\pm$   0.8 &          13.7 $\pm$   1.1 &          11.2 $\pm$   1.6 &           9.9 $\pm$   3.6 \\
     
      XTE~J1709-267   &   357.48 &      7.9 &           5.0 $\pm$   0.5 &           4.5 $\pm$   1.1 &           2.6 $\pm$   2.1 &                 $<$   4.9 \\
     
        GRS~1734-294 &   358.90 &      1.4 &          11.7 $\pm$   0.5 &           7.4 $\pm$   1.0 &           6.1 $\pm$   2.1 &                 $<$   4.7 \\
       Sco~X-1$^{*}$ &   359.09 &     23.8 &         205.1 $\pm$   1.2 &          16.4 $\pm$   1.9 &          11.5 $\pm$   3.8 &                 $<$   8.2 \\
     1E~1740.7-2942  &   359.12 &     -0.1 &          49.6 $\pm$   1.0 &          52.5 $\pm$   1.2 &          48.3 $\pm$   1.6 &          39.2 $\pm$   3.6 \\
        SLX~1744-299 &   359.28 &     -0.9 &          16.9 $\pm$   0.8 &           8.6 $\pm$   2.1 &                 $<$   8.2 &                           \\
       XTE ~J817-330 &   359.82 &     -8.0 &           7.3 $\pm$   0.4 &           6.6 $\pm$   0.9 &           4.7 $\pm$   1.7 &                $<$   7.8 \\
           V2400~Oph &   359.86 &      8.7 &           6.7 $\pm$   0.7 &                 $<$   1.7 &                           &                           \\
         Oph~Cluster &     0.59 &      9.3 &           4.4 $\pm$   0.7 &           2.6 $\pm$   1.7 &                 $<$   3.2 &                           \\
   
%     IGR~J17475-2822  &     0.61 &     -0.1 &          32.9 $\pm$   0.5 &          20.2 $\pm$   1.1 &           9.2 $\pm$   2.3 &          12.0 $\pm$   5.4 \\
     IGR~J17475-2822  &     0.61 &     -0.1 &          32.9 $\pm$   0.5 &          20.2 $\pm$   1.1 &               $<$   4.6 &                          \\

        SLX~1735-269 &     0.79 &      2.4 &          12.4 $\pm$   0.4 &           7.7 $\pm$   0.9 &                 $<$   1.8 &                           \\
        RX~J1832-330 &     1.53 &    -11.4 &           8.9 $\pm$   0.5 &           6.7 $\pm$   1.0 &           7.7 $\pm$   2.0 &              $<$   9.0 \\
              GX~1+4 &     1.93 &      4.8 &          48.3 $\pm$   0.4 &          32.8 $\pm$   0.9 &           6.2 $\pm$   1.7 &                 $<$   3.5 \\
              GX~3+1 &     2.30 &      0.8 &           4.6 $\pm$   0.5 &           2.8 $\pm$   1.1 &                 $<$   2.3 &                           \\
         4U~1820-303 &     2.78 &     -7.9 &          11.6 $\pm$   0.4 &           1.4 $\pm$   1.0 &                 $<$   4.0 &                           \\
        GRS~1758-258 &     4.50 &     -1.4 &          62.3 $\pm$   0.7 &          79.2 $\pm$   1.7 &          83.4 $\pm$   1.6 &          47.4 $\pm$   3.6 \\

         V1223~Sgr  &     4.97 &    -14.3 &           5.8 $\pm$   0.6 &                 $<$   1.5 &                           &                           \\
              GX~5-1 &     5.08 &     -1.0 &          20.4 $\pm$   0.8 &           4.2 $\pm$   1.7 &                 $<$   3.4 &                           \\
           V2487~Oph &     6.59 &      7.8 &           2.1 $\pm$   0.5 &                 $<$   1.1 &                           &                           \\
     IGR~J18173-2509 &     6.78 &     -4.3 &           3.1 $\pm$   0.4 &           3.2 $\pm$   0.9 &                 $<$   3.4 &            \\
     IGR~J17597-2201 &     7.56 &      0.8 &           3.3 $\pm$   0.8 &           4.1 $\pm$   1.9 &           5.1 $\pm$   3.6 &                           \\
     IGR~J17586-2129 &     8.05 &      1.3 &           4.1 $\pm$   0.9 &           2.5 $\pm$   2.2 &                 $<$   4.2 &                           \\
   
     IGR~J17513-2011 &     8.17 &      3.4 &           2.8 $\pm$   0.4 &           3.1 $\pm$   1.0 &                 $<$   2.0 &                           \\
    
              GX~9+9 &     8.50 &      9.0 &           4.0 $\pm$   0.5 &                 $<$   1.2 &                           &                           \\
          GS~1826-24 &     9.27 &     -6.1 &          83.1 $\pm$   0.4 &          66.1 $\pm$   1.0 &          33.6 $\pm$   1.9 &          24.5 $\pm$   3.9 \\
     SAX~J1802.7-201 &     9.42 &      1.0 &           8.1 $\pm$   0.6 &                 $<$   1.6 &                           &                           \\
         SGR~1806-20 &     9.98 &     -0.2 &           5.4 $\pm$   0.6 &           4.0 $\pm$   1.2 &                 $<$   2.4 &                           \\
    
      PSR~J1811-1926$^{b}$ &    11.17 &     -0.3 &           3.7 $\pm$   0.6 &           7.7 $\pm$   1.2 &           6.3 $\pm$   1.8 &             $<$   8.0 \\
   
   HETE~J1900.1-2455 &    11.30 &    -12.9 &          15.8 $\pm$   0.7 &          10.9 $\pm$   1.5 &           7.1 $\pm$   2.8 &                 $<$   6.0 \\
        PKS~1830-211 &    12.17 &     -5.7 &           3.8 $\pm$   0.5 &           5.4 $\pm$   1.1 &           4.3 $\pm$   2.1 &                 $<$   9.4 \\
  
   IGR~J18135-1751$^{+}$   &    12.79 &      0.0 &           4.6 $\pm$   0.9 &           4.4 $\pm$   2.3 &                 $<$   4.5 &                           \\

             GX~13+1 &    13.52 &      0.1 &           8.1 $\pm$   0.9 &           6.6 $\pm$   2.2 &                 $<$   4.3 &                           \\
      
      2E~1739.1-1210 &    13.93 &      9.4 &           4.5 $\pm$   0.6 &           6.6 $\pm$   1.3 &                 $<$   4.8 &                           \\
      
            NGC~7172 &    15.12 &    -53.1 &           6.9 $\pm$   1.6 &          12.1 $\pm$   4.4 &                 $<$   8.6 &                           \\
             GX~17+2 &    16.43 &      1.3 &          20.6 $\pm$   0.8 &                 $<$   1.9 &                           &                           \\
     AX~J1820.5-1434 &    16.48 &      0.1 &           3.5 $\pm$   0.8 &           4.7 $\pm$   1.9 &                 $<$   7.4 &                           \\
     IGR~J18214-1318 &    17.69 &      0.5 &           3.8 $\pm$   0.7 &           2.6 $\pm$   1.9 &                 $<$   3.6 &                           \\
           M~1812-12 &    18.03 &      2.4 &          25.9 $\pm$   0.6 &          24.2 $\pm$   1.3 &          12.9 $\pm$   2.4 &                 $<$   5.0 \\
    IGR~J183047-1232 &    19.44 &     -1.2 &           3.2 $\pm$   0.6 &           3.4 $\pm$   1.5 &                 $<$   5.8 &                           \\
      SNR~021.5-00.9 &    21.51 &     -0.9 &           4.9 $\pm$   1.0 &           3.4 $\pm$   2.5 &                 $<$   4.8 &                           \\
     
     AX~J183039-1002$^{+}$ &    21.67 &      0.0 &           4.4 $\pm$   1.0 &           3.7 $\pm$   2.5 &                 $<$   4.8 &                           \\
     
     IGR~J18325-0756 &    23.71 &      0.6 &           3.5 $\pm$   0.7 &           4.8 $\pm$   1.8 &           4.9 $\pm$   3.4 &                           \\
  
  SWIFT~J1753.5-0127$^{*}$ &    24.90 &     12.2 &          12.4 $\pm$   1.3 &          18.4 $\pm$   2.0 &          14.5 $\pm$   3.5 &                 $<$   7.6 \\
  
   AX~J1838.0-0655  &    25.26 &     -0.2 &           4.4 $\pm$   0.7 &           3.9 $\pm$   1.9 &                 $<$   3.7 &                           \\

         4U~1850-087 &    25.35 &     -4.3 &           8.7 $\pm$   0.6 &           4.1 $\pm$   1.5 &                 $<$   5.8 &                           \\
   
     AX~J1841.0-0535$^{+}$ &    26.77 &     -0.2 &           7.9 $\pm$   0.7 &          10.2 $\pm$   1.3 &          11.5 $\pm$   2.5 &                 $<$  11.0 \\
     RX~J1940.1-1025$^{+}$ &    28.98 &    -15.5 &           9.8 $\pm$   1.3 &           8.3 $\pm$   2.7 &                 $<$   9.4 &                           \\
     IGR~J18483-0311$^{+}$ &    29.75 &     -0.7 &           7.6 $\pm$   1.0 &                 $<$   5.2 &                           &                           \\
      
          4U1822-000 &    29.94 &      5.8 &           2.4 $\pm$   0.7 &                 $<$   1.5 &                           &                           \\
         3A~1845-024 &    30.42 &     -0.4 &           4.4 $\pm$   1.1 &                  $<$  5.8 &                           &                           \\
       XTE~J1855-026 &    31.08 &     -2.1 &          11.4 $\pm$   0.6 &           6.8 $\pm$   1.3 &           2.9 $\pm$   2.5 &                 $<$   5.7 \\
         4U~1916-053 &    31.35 &     -8.5 &           7.2 $\pm$   0.7 &           7.6 $\pm$   1.5 &           8.8 $\pm$   2.9 &                 $<$   6.4 \\
    
     IGR~J18485-0047  &    31.90 &      0.3 &           6.2 $\pm$   0.7 &           7.0 $\pm$   1.2 &           2.5 $\pm$   2.4 &                 $<$   5.5 \\
    
         GS~1843+009 &    33.05 &      1.7 &           4.7 $\pm$   0.6 &           2.2 $\pm$   1.4 &                 $<$   5.4 &                           \\
       XTE~J1901+014 &    35.37 &     -1.6 &           4.3 $\pm$   0.6 &           3.0 $\pm$   1.4 &                 $<$   2.6 &                           \\
       Aql~X-1$^{*}$ &    35.71 &     -4.1 &          16.6 $\pm$   0.8 &          13.9 $\pm$   1.2 &           5.8 $\pm$   2.3 &                 $<$   5.3 \\
             Ser~X-1 &    36.12 &      4.8 &           3.1 $\pm$   0.5 &           1.5 $\pm$   1.3 &                 $<$   2.5 &                           \\
       XTE~J1858+034 &    36.82 &     -0.1 &          10.3 $\pm$   0.7 &           4.3 $\pm$   1.6 &                 $<$   6.4 &                           \\
          4U~1901+03 &    37.18 &     -1.3 &          14.1 $\pm$   0.7 &                 $<$   1.7 &                           &                           \\
              SS~433 &    39.69 &     -2.2 &           6.5 $\pm$   0.5 &           2.7 $\pm$   1.2 &                 $<$   5.6 &                          \\

          4U~1909+07 &    41.90 &     -0.8 &          11.1 $\pm$   0.5 &           4.7 $\pm$   1.3 &                 $<$   5.0 &                           \\
      
       XTE~J1908+094$^{+}$ &    43.27 &      0.4 &          14.1 $\pm$   0.7 &           2.7 $\pm$   1.6 &                 $<$   6.4 &                           \\
   
   IGR~J19140+0951   &    44.29 &     -0.5 &           8.6 $\pm$   0.9 &          11.4 $\pm$   1.7 &                 $<$   3.3 &                           \\
  GRS~1915+105$^{*}$ &    45.36 &     -0.2 &         207.0 $\pm$   1.0 &          93.1 $\pm$   1.6 &          61.8 $\pm$   2.3 &          38.6 $\pm$   5.2 \\
      2E~1853.7+1534 &    47.40 &      6.1 &           2.5 $\pm$   0.6 &           3.1 $\pm$   1.4 &                 $<$   2.8 &                           \\
    
     IGR~J19443+2117  &    57.79 &     -1.4 &           6.2 $\pm$   0.9 &           2.8 $\pm$   2.1 &                 $<$   3.9 &                           \\
    
             Her~X-1 &    58.15 &     37.5 &          44.4 $\pm$   1.6 &          10.5 $\pm$   4.3 &                $<$   17.0 &                           \\
             
               3C382 &    61.32 &     17.4 &          21.1 $\pm$   3.5 &          11.2 $\pm$   7.2 &                 $<$  11.3 &                           \\
         KS~1947+300 &    66.10 &      2.1 &          18.9 $\pm$   0.9 &           9.1 $\pm$   2.2 &          12.8 $\pm$   4.0 &                $<$   17.6 \\
      Cyg~X-1$^{**}$ &    71.33 &      3.1 &         811.0 $\pm$   1.1 &         876.5 $\pm$   1.7 &         769.9 $\pm$   3.1 &         500.6 $\pm$   6.9 \\
               Cyg~A &    76.21 &      5.8 &           2.8 $\pm$   0.6 &                 $<$   1.6 &                           &                           \\
        EXO~2030+375 &    77.14 &     -1.2 &          39.0 $\pm$   0.6 &          15.4 $\pm$   1.5 &                 $<$   2.8 &                           \\
             Cyg~X-3 &    79.84 &      0.7 &         146.2 $\pm$   0.6 &          56.2 $\pm$   1.5 &          25.0 $\pm$   2.9 &          14.8 $\pm$   6.2 \\

    SAX~J2103.5+4545 &    87.12 &     -0.7 &          11.7 $\pm$   0.6 &                 $<$   1.6 &                           &                           \\
             Cyg~X-2 &    87.32 &    -11.3 &           8.8 $\pm$   0.8 &                 $<$   2.1 &                           &                           \\
   
     IGR~J21335+5105  &    94.41 &     -0.5 &          11.0 $\pm$   0.8 &                 $<$   2.2 &                           &                           \\
   
         4U~2206+543 &   100.60 &     -1.1 &           4.1 $\pm$   0.8 &           5.1 $\pm$   2.1 &          11.6 $\pm$   3.7 &                 $<$   7.6 \\
               Cas~A &   111.71 &     -2.1 &           2.6 $\pm$   0.4 &           3.0 $\pm$   1.1 &                 $<$   2.2 &                           \\
             709~Cas &   120.07 &     -3.4 &           3.3 $\pm$   0.4 &                 $<$   1.2 &                           &                           \\
       $\gamma$~Cas &   123.55 &     -2.2 &           3.0 $\pm$   0.5 &           1.3 $\pm$   1.3 &                 $<$   2.5 &                           \\
         1A~0114+650 &   125.71 &      2.6 &           6.7 $\pm$   0.6 &           2.7 $\pm$   1.6 &                 $<$   3.1 &                           \\
            X0115+63 &   125.92 &      1.0 &          19.8 $\pm$   0.9 &           8.3 $\pm$   1.3 &                 $<$   5.2 &             \\
           4U0142+61 &   129.38 &     -0.4 &           3.3 $\pm$   0.7 &           2.9 $\pm$   1.7 &                 $<$   3.0 &                           \\
         GT~0236+610 &   135.68 &      1.1 &           3.6 $\pm$   0.9 &           4.8 $\pm$   2.4 &                 $<$   4.5 &                           \\
                MKN3 &   143.30 &     22.7 &           5.9 $\pm$   1.3 &           7.4 $\pm$   3.7 &                 $<$   7.0 &                           \\
      V0332+53$^{*}$ &   146.05 &     -2.2 &         105.4 $\pm$   1.9 &          11.7 $\pm$   3.1 &                 $<$   6.0 &                           \\
            NGC~4151 &   155.08 &     75.1 &          28.2 $\pm$   1.4 &          34.4 $\pm$   3.9 &          22.9 $\pm$   6.8 &          46.9 $\pm$  14.1 \\
              X~Per &   163.09 &    -17.1 &          22.7 $\pm$   2.6 &          41.1 $\pm$   6.2 &          27.7 $\pm$  11.2 &                 $<$  23.9 \\
    
             NGC1068  &   172.14 &    -51.9 &           4.7 $\pm$   1.2 &           9.9 $\pm$   3.3 &           7.7 $\pm$   6.5 &

\enddata

\tablecomments{The catalog contains only the sources detected
in the 50-150 keV band. The counts for diffuse continuum and sources 
were extracted simultaneously to obtain these data.(**) Source tentatively identified.} 
(*) are sources variables on pointing time scale ($\sim$ 2500 s in the 25-50 keV band. Cyg X-1(***) is variable 
up to 200 keV. (+) more than one possible identification
(a) Sources are separated in the 20-50 keV band, but may be confused at higher energies.\\
(b) can be affected by SGR 1806-20.
 
\end{deluxetable}

%%%%%%%%%%%%%%%%%%%%%%%%%%%%%%%%%%%%%%%%%%%%%%%%%%%%%%%%%%%%%%%%%%%%%%%%%%%%%
%%%% Table 2 %%%%%%%%%%%%%%%%%%%%%%%%%%%%%%%%%%%%%%%%%%%%%%%%%%%%%%%%%%%%%%%%
%%%%%%%%%%%%%%%%%%%%%%%%%%%%%%%%%%%%%%%%%%%%%%%%%%%%%%%%%%%%%%%%%%%%%%%%%%%%%
\begin{deluxetable}{lccccccccc}%1
\tablewidth{0pt}
\tablecaption{The best fit  parameters for the spectra presented on fig~\ref{fig:spectrum}}

\tabletypesize{\scriptsize}
\tablehead{
 \colhead{}\\
%\colhead{}
%&\colhead{low energy}  
%&\colhead{Cutoff power law} 
% &\colhead{}
%&\colhead{power }  
%&\colhead{  law} 
% &\colhead{$F_{511 keV}$} 
%%&\colhead{~}         
%&\colhead{$F_{posit}$} 
%&\colhead{$f_{p}$} \\
\colhead{ }
&\colhead{index1} 
&\colhead{Ecut} 
&\colhead{$F1_{50 keV}$}
&\colhead{index2} 
&\colhead{$F2_{50 keV}$}
&\colhead{$F_{511 keV}$} 
%%&\colhead{~}         
&\colhead{$F_{posit}$} 
&\colhead{$f_{p}$} \\
 &&&\colhead{$\times 10^{-4}~ph$}
  &&\colhead{$\times 10^{-4}~ph~$}
 &\colhead{$\times 10^{-4}~ph~$}
 &\colhead{$\times 10^{-4}~ph~$}\\
 %&\colhead{}

 &&\colhead{keV }

&\colhead{$~cm^{-2}~s^{-1}~keV^{-1}$}

&\colhead{}

&\colhead{$cm^{-2}~s^{-1}~keV^{-1}$}

&\colhead{$ cm^{-2}~s^{-1}$}
&\colhead{$ cm^{-2}~s^{-1}$}
 
 }
\startdata

     Diffuse 1&   0 (fixed) &  7.5  $\pm$ 1 & 0.66 $\pm$ 0.05 &   &    &          &    &    \\
     Diffuse 2      &    &       &  & 1.55 $\pm$   0.25 &  1.4  $\pm$   0.15  &                        &    &    \\
     Diffuse 3&   &   &   &   &     &  8.68 $\pm$   0.61             &  36.1 $\pm$ 4.2 & 0.98 $\pm$ 0.05   \\
    Sources &    2.67   $\pm$ 0.04     &   &  24.7$\pm$ 1.0   &       &         &        &  & \\
  
\enddata

\tablecomments{Diffuse 1 corresponds to the component at low energy, potentially associated with a 
white dwarf population.\\
Diffuse 2 corresponds to the "true" diffuse emission, potentially related to the CR particles.\\
Diffuse 3 corresponds to the annihilation processus. \\
Sources :  a 5\% systematic error is added to the combined spectrum.
}
 
\end{deluxetable}

\end{document}